\documentclass[reprint, aps, prapplied, superscriptaddress, amsmath, amssymb]{revtex4-2}

\usepackage[utf8]{inputenc}
\usepackage[T1]{fontenc}
\usepackage{hyperref}
\usepackage{graphicx}
\usepackage{verbatim}

\begin{document}

\title{A Voltage-Controlled Josephson Frequency Comb}

\author{Giorgio De Simoni}
\email{giorgio.desimoni@nano.cnr.it}
\affiliation{NEST, Istituto Nanoscienze-CNR and Scuola Normale Superiore, Piazza S. Silvestro 12, I-56127 Pisa, Italy}

\author{Francesco Giazotto}
\affiliation{NEST, Istituto Nanoscienze-CNR and Scuola Normale Superiore, Piazza S. Silvestro 12, I-56127 Pisa, Italy}


\begin{abstract}
Microwave frequency combs constitute promising resources for quantum technologies, cryogenic electronics, and multiplexed sensing architectures. In this work, we propose a frequency-comb generator based on a Josephson field-effect transistor operated in a relaxation-oscillation regime. The device comprises a gate-tunable ballistic superconductor–semiconductor–superconductor junction embedded in a resistively shunted circuit, in which electrostatic control of the carrier density enables in situ tuning of both the critical current and the Josephson inductance. Time-domain circuit simulations indicate that the resulting oscillator produces coherent voltage pulses whose Fourier spectrum forms a microwave frequency comb. In contrast to conventional Josephson-based comb architectures, the proposed platform provides direct electrical control of the comb spacing, emission frequencies, and modal power distribution via a gate electrode. For a representative Al/InAs implementation, we demonstrate continuous frequency coverage in the technologically relevant 1–10 GHz range. Furthermore, the concept is shown to be compatible with higher-$T_c$ superconductors, underscoring its potential as a compact and scalable microwave source for cryogenic quantum information and sensing applications.
\end{abstract}

\maketitle

\section{Introduction}

A frequency comb is an electromagnetic field composed of a series of phase-coherent, regularly spaced spectral modes that, in the time domain, correspond to a periodic train of ultrashort pulses. Over the past several decades, interest in optical frequency comb sources has grown substantially, motivated by their capacity to provide unprecedented control over the frequencies of emitted photons \cite{udem2002optical}. This control, typically achieved by generating higher-order harmonics from a fundamental pump field, has enabled the development of a diverse range of radiation-generation schemes \cite{burghoff2015, hillbrand2020prl}. These developments have significantly broadened the accessible bandwidth of frequency combs, thereby driving transformative advances in precision metrology \cite{hansch1999laser}, high-resolution spectroscopy \cite{RevModPhys.54.685, hansch1994frontiers}, next-generation telecommunication infrastructures \cite{foreman2007remote}, and quantum information science and technology \cite{reviewcomb_optics}.

\begin{figure}[ht!]
\includegraphics[width=\linewidth]{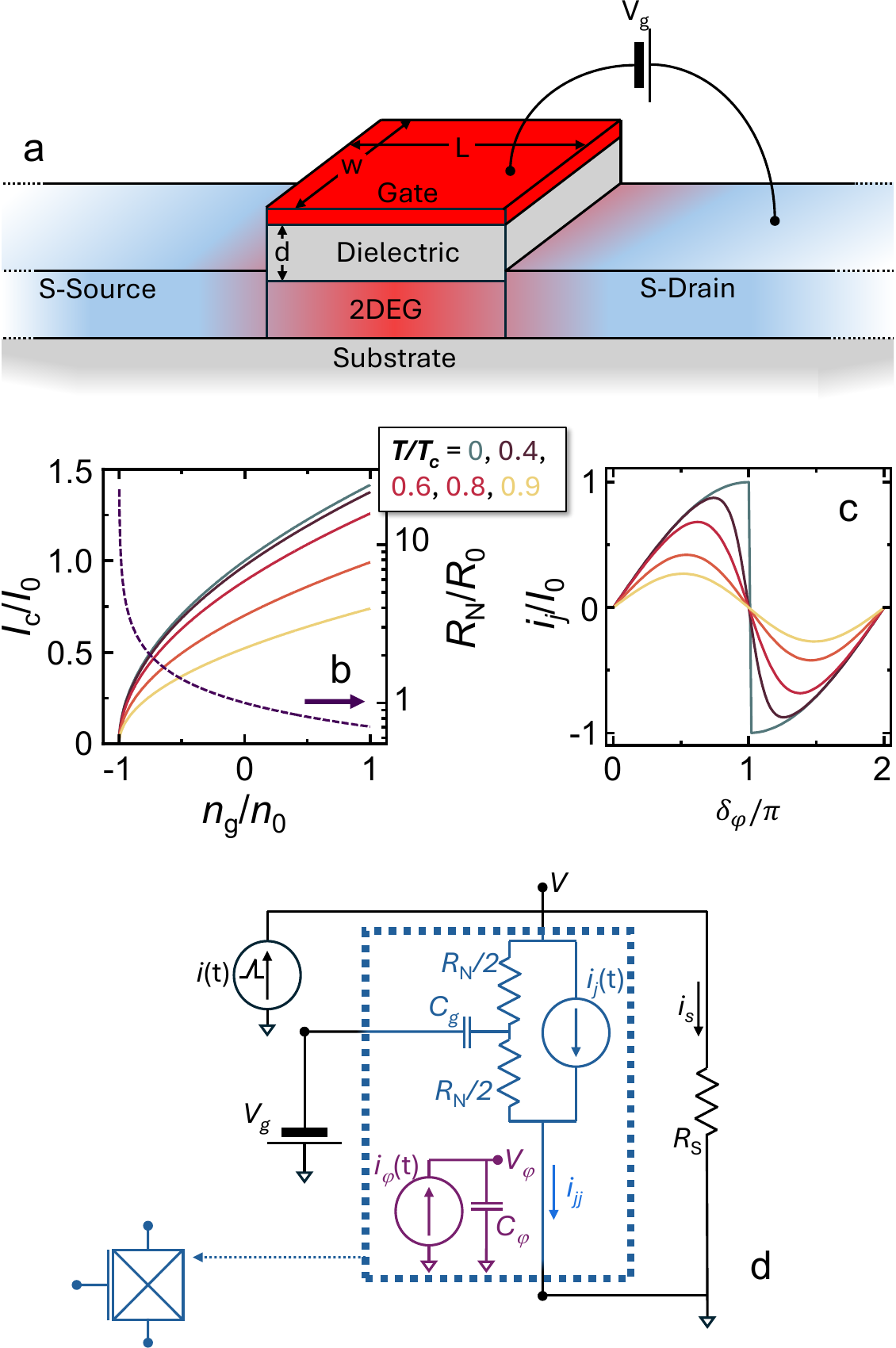}
\caption{\label{fig:fig1} 
(a) Schematic representation of a gate-tunable superconductor–normal-metal–superconductor (SNS) Josephson junction with length \(L\) and width \(W\). The device consists of two superconducting electrodes (light blue) that contact a semiconductor region hosting a two-dimensional electron gas (2DEG, red), which is capacitively coupled to a gate electrode (red) through a dielectric layer (grey). (b) Dependence of the normalized critical current \(I_c/I_0\) (left axis) and of the normal-state resistance \(R_N\) (right axis), expressed in units of \(R_0 \equiv \frac{h}{2e^2 W n^2}\), on the induced carrier density \(n_g/n_0\) for selected temperatures (given in units of \(T_c\)). (c) Supercurrent–phase relation (C\(\Phi\)R) obtained from Eq.~(\ref{eq:CphiR}) and displayed for the same set of temperatures as in panel (b). (d) Equivalent RSJ circuit of the SNS junction (blue), comprising the normal-state resistance and the Josephson branch characterized by the junction C\(\Phi\)R, together with the capacitive coupling to the gate electrode and the auxiliary phase-dynamics loop (purple) used to implement the second Josephson relation within the SPICE solver. \(i(t)\), \(i_j(t)\), \(i_{jj}(t)\), \(i_s(t)\), and \(i_\varphi(t)\) denote the bias current, the Josephson current, the total current flowing through the SNS junction, the current through the shunt resistor, and the virtual current charging the phase capacitor, respectively. The circuit symbol representing the gate-controlled SNS Josephson junction is also shown.
}
\end{figure}

The demonstrated efficacy of these optical platforms has, in recent years, motivated analogous investigations within condensed-matter physics. These efforts focus on the realization of compact, versatile solid-state architectures designed to mitigate the substantial technological gap arising from the absence of tunable, cryogenic sources of broadband microwave frequency combs. This objective is of particular importance for superconducting and spin-based qubits, which generally require sub-\(8\,\mathrm{GHz}\) microwave signals for dispersive readout and coherent control \cite{martinis_Xmon, crippa2019gate}.
Within this framework, Josephson-junction-based oscillators \cite{koshelets_integrated_2000, galin_towards_2015} constitute a leading technological platform. Theoretical proposals indicate that Josephson radiation comb generators (JRCGs) can be realized either in extended Josephson junctions \cite{solinas2015extended} or in superconducting quantum interference devices (SQUIDs) driven by time-dependent magnetic fields \cite{solinas2015jrcg}. These mesoscopic quantum devices can function as efficient multitone synthesizers \cite{JAWS, fernando_multiplier} and as sources of coherent microwave photons \cite{lahteenmaki2013, Mottonen2021}, exhibiting lasing-like emission \cite{astafiev2007, cassidy, liu2015semiconductor} and enabling microscopic implementations in the framework of quantum optics \cite{nori_review}. Although very recent experiments have demonstrated coherent microwave frequency-comb generation in SQUID-based architectures \cite{greco2025coherent} and their application to multiplexed spectroscopy \cite{greco2026spectroscopy}, most existing realizations still encounter significant limitations. In particular, many implementations employ long chains of Josephson junctions \cite{IEEE_2020} to compensate for intrinsically low power output \cite{464849, galin_towards_2015}, thereby introducing scalability challenges, or they rely on cavity-embedded architectures \cite{Pappas_2014, wang2021, shin2022, wang2024, acoustic}, which typically offer only restricted in-situ tunability of the generated bandwidth and impose stringent constraints on repetition frequency \cite{bao2024}.

A different category of superconducting oscillators is constituted by relaxation oscillators \cite{toomey_frequency_2018, toomey_microwave_2017} (SROs). These devices are intrinsically nonlinear dynamical systems that exploit the rapid switching between the zero-resistance superconducting state and the dissipative normal state of a superconducting element. Common implementations employ either a superconducting nanowire or a Josephson junction shunted by an inductive–resistive network, so that the operation relies on the hysteretic behavior intrinsic to the current–voltage ($I$–$V$) characteristics.
In an SRO biased at a constant current $i$ that exceeds the critical current $I_c$, the current initially flows through the superconducting branch. Once the current exceeds $I_c$, the branch undergoes an abrupt transition to the normal state, producing a voltage spike and redirecting the current into the lower-resistance shunt branch. The system then enters a relaxation step during which the current through the superconducting element decreases. When this current falls below $I_c$, the superconducting state is reinstated, and the oscillatory cycle restart. The oscillation frequency is predominantly determined by the characteristic time constant $\tau \approx L/R$, where $L$ and $R$ denote the effective inductance and resistance of the oscillator, respectively.
In practical circuits, effective inductance $L$ arises from both the intrinsic reactive response of the superconducting device and the deliberate or parasitic inductances of the surrounding circuitry. Similarly, the effective resistance $R$ is determined by the normal-state resistance $R_N$ of the superconducting element, as well as other dissipative components in the SRO. Due to their capability to generate sub-nanosecond voltage pulses with low energy dissipation, these oscillators are key building blocks in Single Flux Quantum (SFQ) digital logic circuits \cite{likharev1991} and serve as biomimetic spiking primitives in superconducting neuromorphic computing architectures \cite{toomey_design_2019}.
The electrical and spectral properties of SROs are frequently tuned through controlled local heating of the device \cite{mccaughan_superconducting-nanowire_2014, baghdadi_multilayered_2020, PhysRevB.79.100509, Trupiano2024QuasiparticleInjection}, which provides a powerful control parameter. However, such thermal modulation is less desirable in quantum electronic architectures, where thermal noise-induced decoherence imposes stringent constraints on allowable dissipation and temperature fluctuations.

In this work, we establish a connection between these concepts by introducing a microwave frequency-comb generator based on a Josephson field-effect transistor operated as a relaxation oscillator (JRO). The device consists of a resistively-shunted superconductor–normal metal–superconductor (SNS) Josephson junction, in which the normal (N) region is implemented using a two-dimensional electron gas (2DEG) confined in the quantum well of a semiconductor heterostructure. In this configuration, the inherently periodic output voltage of the relaxation oscillator gives rise to a phase-coherent frequency-comb spectrum. 
A key feature of this architecture is that the semiconducting 2DEG enables field-effect modulation of the carrier density via a gate electrode. This capability offers a straightforward and continuous means of tuning the junction critical current, which affords precise in situ control over the fundamental oscillation frequency, the emitted power, and the overall bandwidth of the generated comb. Consequently, this platform integrates the compact footprint characteristic of mesoscopic Josephson junctions with the high degree of tunability required for advanced cryogenic electronics and emerging quantum technologies.

\section{A gate-controlled ballistic Josephson Junction}
The proposed device architecture is schematically depicted in \autoref{fig:fig1}a. Two superconducting electrodes (S, light blue in \autoref{fig:fig1}) are fabricated to form high-quality electrical contact with a semiconductor region that hosts a two-dimensional electron gas (2DEG, red). 
Structures of this type have been realized using a variety of material platforms and fabrication strategies, including semiconductor nanowires \cite{doi:10.1126/science.1113523,PhysRevB.89.214508,Paajaste2015,Kousar2022,roddaro2011hot,spathis2011hybrid,strambini2020josephson,iorio2018vectorial}, exfoliated or transferred semiconductor flakes \cite{Heersche2007}, modulation-doped quantum wells \cite{10.1063/1.103546}, and surface charge accumulation layers \cite{PhysRevB.106.235404,Paghi2025,10.1063/5.0225361,Paghi2025a,Paghi2025b}. 
In most reported implementations, 2DEG is formed in InAs or In$_x$Ga$_{1-x}$As with $x\geq 0.75$ \cite{capotondi2004two,desrat2004magnetotransport}, mainly because the negligible Schottky barrier at the superconductor–semiconductor interface facilitates efficient Andreev reflection, thereby improving superconducting correlations across the junction \cite{10.1063/1.95809,10.1063/1.119116,Mayer2020,amado2013electrostatic,amado2014ballistic}. However, gate-tunable Josephson junctions based on alternative two-dimensional material systems, most prominently graphene, have also been demonstrated \cite{Heersche2007}. When the length $L$ of the S–Semiconductor–S weak link is sufficiently short, the structure supports a dissipationless supercurrent mediated by the superconducting proximity effect, which arises from the formation of Andreev bound states in the semiconductor region \cite{Andreev1964,Pannetier2000,RevModPhys.76.411}.

In this work, we consider a junction that operates in the ballistic transport regime and within the short-junction approximation. This condition is satisfied in the short-junction limit, in which the junction length \(L\) is much smaller than both the superconducting coherence length in the normal region, \(\xi_N = \hbar v_F / 2\pi k_B T_c\), and the electronic mean free path \(l_N\) in the normal metal, with \(v_F\) denoting the Fermi velocity of the two-dimensional electron gas.
In this regime, following the derivation shown in \cite{PhysRevLett.66.3056}, $I_c$ can be expressed as
\begin{equation}
I_c(T)=Ne\frac{\Delta(T)}{\hbar},
\end{equation}
where $\Delta(T)\simeq1.764 k_B T_C\tanh\left(1.74\sqrt{\frac{Tc}{T}-1}\right)$ is the temperature-dependent energy gap of the S electrodes with the critical temperature $T_c$, and $k_B$ is the Boltzmann constant. For a two-dimensional system,
\begin{equation}
N=\frac{W}{\lambda_f}\equiv 2 \pi W \sqrt{2 \pi n}
\end{equation}
is the number of ballistic channels in the junction, which is equivalent to the ratio of the junction width $W$ to the Fermi wavelength $\lambda_f$. The latter exhibits a square-root dependence on the carrier density $n$, which can conveniently be exploited to tune both $I_c$ and the normal-state resistance $R_N$ of the junction through the application of a control voltage ($V_g$) to a gate electrode (colored in red in \autoref{fig:fig1}), electrically insulated and capacitively coupled to the 2DEG by a dielectric layer (colored in gray in \autoref{fig:fig1}) of thickness $d$ and dielectric constant $\epsilon\equiv\epsilon_0\epsilon_r$ (with $\epsilon_0$ and $\epsilon_r$ being the vacuum and relative dielectric constants, respectively). Finally, $R_N$ is simply given by the inverse of the conductance quantum multiplied by the number of ballistic channels of the junction:
\begin{equation}
\frac{1}{R_N}=N \frac{2e^2}{h}.
\end{equation}
By adopting an approach similar to that shown in \cite{8915971}, $n$ can be written as
\begin{equation}
n=n_0+n_g,
\end{equation}
where $n_0$ is the two-dimensional charge density of the 2DEG at zero gate voltage, and
\begin{equation}
n_g=\frac{C_gV_g}{e}=\frac{\epsilon W L V_g}{d}  
\end{equation}
is the charge accumulated or depleted due to the application of a gate voltage. \autoref{fig:fig1}b (left axis) shows the evolution of the critical current, normalized to $I_0=I_c(n_g=0,T=0)$, as a function of $n_g/n_0$ for a ballistic junction, as described above, at selected temperatures between 0 and 1 in units of $T_c$. In an ideal system, electrostatic control of $n$ enables modulation of $I_c$ up to complete suppression, which also implies a divergence of $R_N$ in $n_g=-n_0$, as shown in \autoref{fig:fig1}b (right axis).

\begin{figure}[t!]
\includegraphics[width=\linewidth]{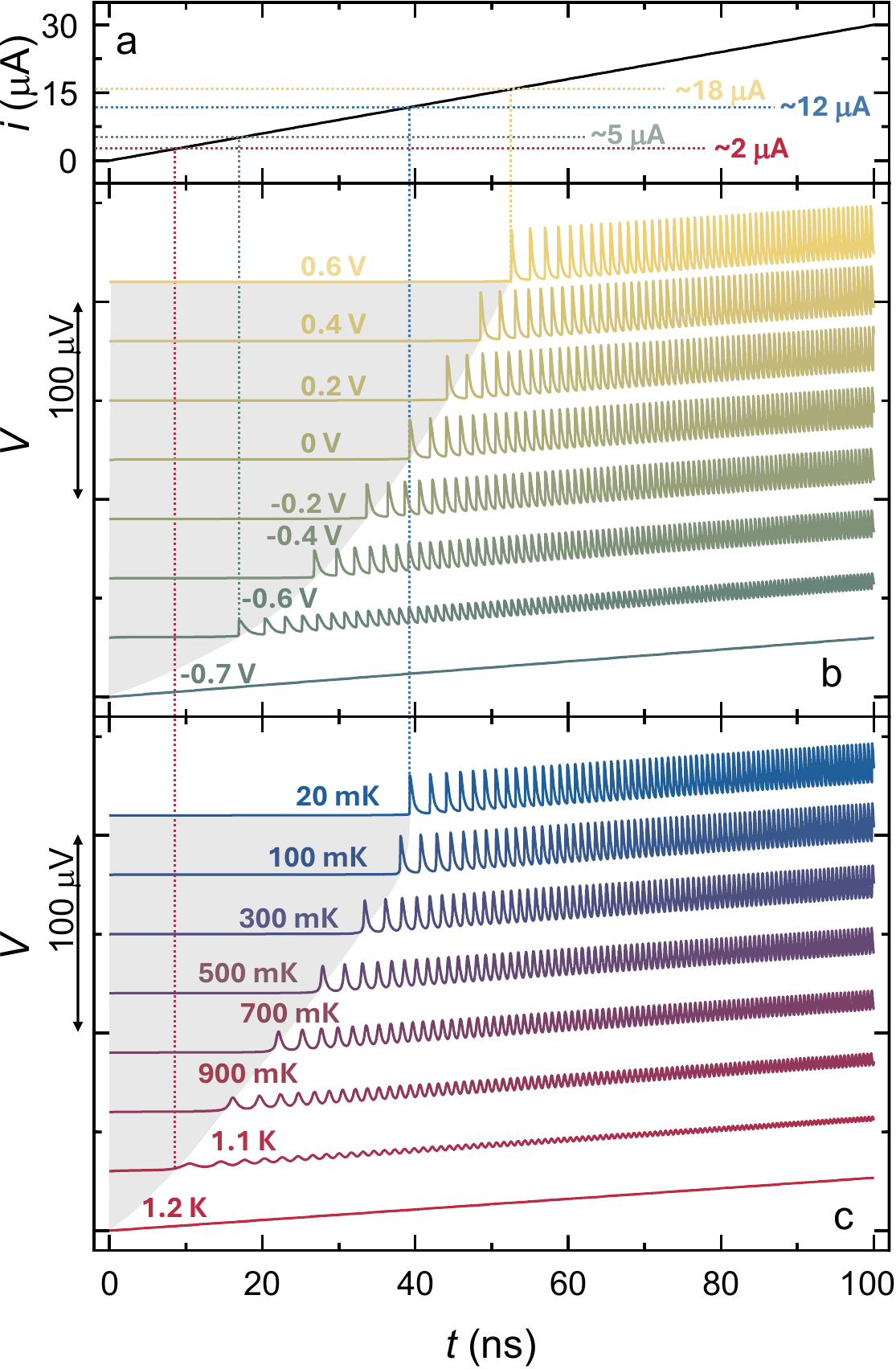}
\caption{\label{fig:fig2} 
Time-dependent response of a shunted, gate-tunable Al–InAs–Al superconductor–normal-metal–superconductor (SNS) ballistic Josephson junction. The traces are obtained from transient numerical simulations with sub-picosecond temporal resolution for representative values of the gate voltage $V_g$ and temperature $T$. (a) Bias current $i$ as a function of time $t$. The current $i$ is increased linearly from 0 to 30 $\mu$A over a time interval of 100 ns. (b) Corresponding voltage characteristics $V(i(t))$ at $T = 20$ mK for gate voltages $V_g$ in the range $-0.7$ to $0.6$ V. A nondissipative regime ($V \approx 0$, gray region) is observed at low bias currents, whereas for $i > I_c(V_g)$ relaxation oscillations emerge, which provides an operational definition of the critical current $I_c$. (c) Voltage characteristics $V(i(t))$ at $V_g = 0$ over the temperature range 20 mK to 1.2 K.
}
\end{figure}

For a fully transparent ballistic junction, the current–phase relation (C$\Phi$R) is described by the Kulik–Omelyanchuk theory in its ballistic limit (KO-2) \cite{RevModPhys.76.411}, which predicts a pronounced non-sinusoidal dependence on the superconducting phase difference $\delta_\varphi$. In this framework, the finite-temperature Josephson current $i_j$ is given by \cite{PhysRevLett.66.3056}
\begin{equation}
\label{eq:CphiR}
i_j(\delta_\varphi)=I_c(T) \sin\left (\frac{\delta_\varphi}{2}\right )\tanh \left[\frac{\Delta(T)}{2k_B T}\cos\left(\frac{\delta_\varphi}{2} \right)\right].
\end{equation}
The corresponding junction C$\Phi$R obtained from \ref{eq:CphiR} is shown in \autoref{fig:fig1}c at the same temperatures as in \autoref{fig:fig1}b. It deviates substantially from that of a conventional Josephson tunnel junction: at low temperatures, it becomes strongly skewed, with its maximum shifted away from $\phi=\pi/2$ and a cusp-like feature emerging near $\phi=\pi$, indicative of the closing of the Andreev minigap. With increasing temperature, thermal broadening progressively smooths the contribution of the Andreev bound states, and the C$\Phi$R continuously evolves towards an approximately sinusoidal dependence.

\section{RSJ and numerical methods}
\begin{figure*}[t!]
\includegraphics[width=\linewidth]{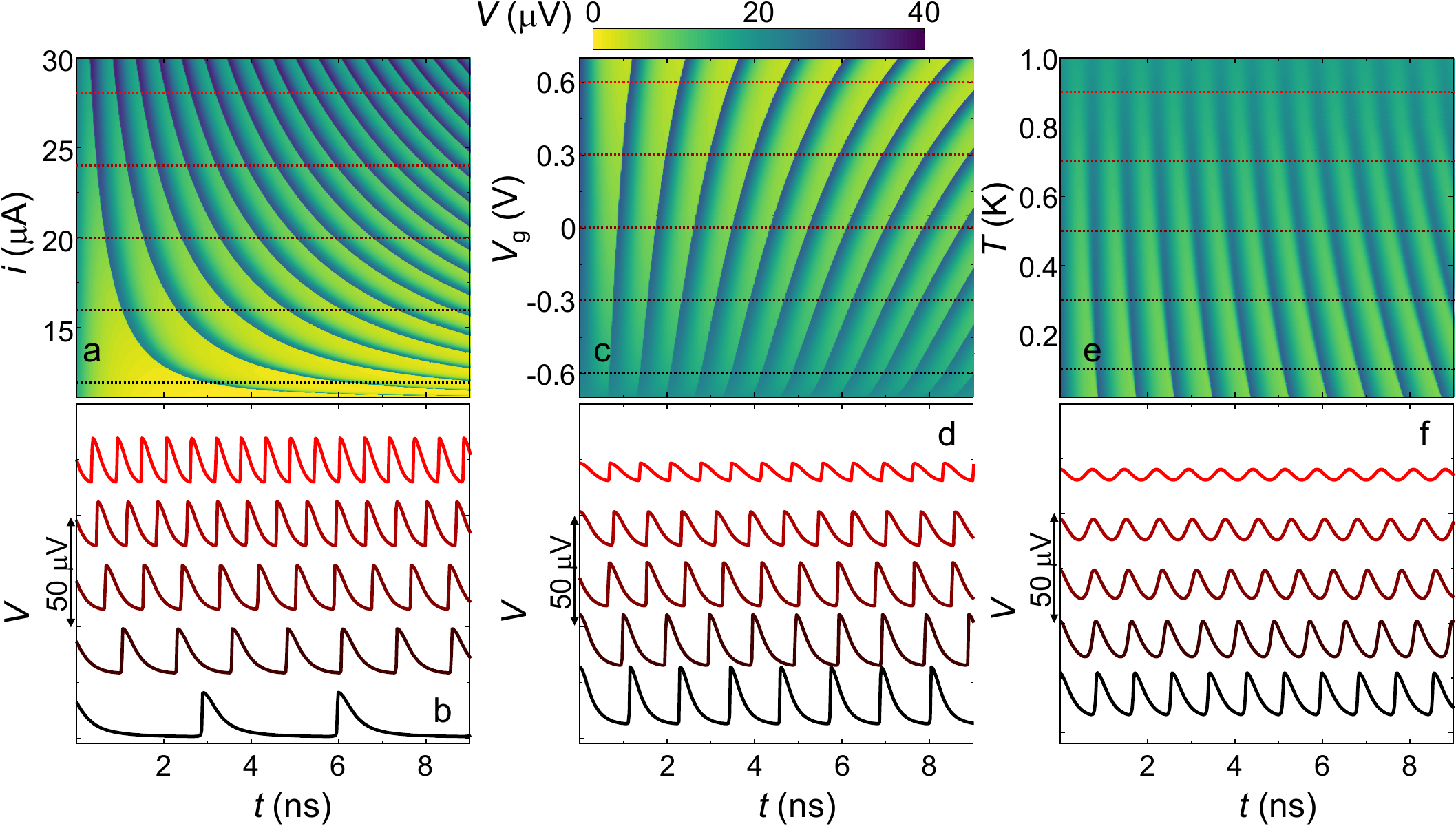}
\caption{\label{fig:fig3}
Time-domain evolution of the JRO output voltage as a function of bias current, gate voltage, and temperature.  
(a) Color map of the output voltage $V(i,t)$ as a function of time and bias current for $T = 20$ mK and $V_g = 0$.  
(b) Representative line cuts extracted from panel (a), illustrating the evolution of the relaxation oscillations with increasing bias current. As $i$ exceeds the critical current $I_c$, the ratio $i/I_c$ increases, reducing the oscillation period $\tau$ from the nanosecond range to a few hundred picoseconds.  
(c) Color map of $V(V_g,t)$ for $T = 20$ mK and fixed $i = 20~\mu$A.  
(d) Selected voltage traces from panel (c), demonstrating the gate tunability of both the oscillation period and amplitude via electrostatic control of the two-dimensional electron gas (2DEG) carrier density and, consequently, of $I_c$. As $V_g$ is swept from depletion ($V_g \sim -0.65$ V) to accumulation ($V_g \sim 0.65$ V), the period $\tau$ decreases by approximately $65\%$, while the oscillation amplitude is reduced by about $40\%$.  
(e) Color map of $V(T,t)$ for fixed $i = 20~\mu$A and $V_g = 0$.  
(f) Representative line cuts from panel (e), showing the temperature dependence of the relaxation dynamics between 20 mK and 1 K. An increase in temperature only weakly modifies the oscillation period (about $15\%$ reduction), but strongly suppresses the voltage amplitude (about $75\%$ reduction). In addition, the oscillation waveform progressively evolves toward a more sinusoidal profile, reflecting thermal modifications of the current–phase relation and the concomitant increase in symmetry between the voltage rise and relaxation phases.}
\end{figure*}

The time-resolved dynamics of a device of the type under consideration can be accurately described within the circuit formalism of the resistively shunted Josephson junction (RSJ) model \cite{Tinkham1996}. In this framework, the Josephson junction is modeled as a parallel combination of a resistor, whose value coincides with the normal-state resistance, and a superconducting branch in which the current $i_j$ is determined by the C$\Phi$R of the junction (\autoref{eq:CphiR}), corresponding to the first Josephson relation in the ballistic short-junction regime. 
It is important to note that, while in conventional tunnel junctions it is generally necessary to include a parallel capacitance to account for the capacitive coupling between the superconducting electrodes, in weak-link-based junctions such as the one investigated here, this capacitive contribution can be neglected, owing to the negligible capacitance between the electrodes. The equivalent circuit of the SNS junction is shown in the blue region of \autoref{fig:fig1}d, where the Josephson branch is represented as an ideal current source. Furthermore, the capacitive coupling between the gate electrode and the parallel network formed by the resistive and Josephson branches is modeled as a capacitor with capacitance $C_g$. 
In our description, this gate capacitance is connected to the circuit at the midpoint of the resistive branch. This choice is an approximation introduced for analytical convenience; it represents one of several possible circuit reductions and does not qualitatively alter the device's overall dynamical behavior.

The Josephson branch must additionally obey the second Josephson relation,
\begin{equation}
\label{eq:Josephson2}
\frac{d\delta_\varphi}{dt} = \frac{2\pi}{\Phi_0} V(t),
\end{equation}
which connects the voltage drop $V(t)$ across the SNS junction to the temporal derivative of the superconducting phase difference. Consequently, solving the time-dependent circuit dynamics involves determining the currents in all branches and the node voltages using Kirchhoff's laws, together with \autoref{eq:CphiR} and \ref{eq:Josephson2}.

The numerical method used in this study is a conventional SPICE-based circuit solver \cite{LTspice}, which is well-suited to simulating the temporal dynamics of Josephson oscillators due to its robust time-domain integration algorithms and its demonstrated capability to handle strongly nonlinear circuits with high numerical stability. The time evolution of the superconducting phase was incorporated into the solver via the second Josephson relation, following the standard approach \cite{Kiviranta2021}, which introduces a fictitious circuit loop to ensure correct computation of the phase difference $\delta_\varphi$.
This approach exploits the equivalence between the integral form of the second Josephson relation
\begin{equation}
\label{eq:Josephson2int}
\delta_\varphi (t)=\frac{2\pi}{\Phi_0}\int_0^t V(t')dt'
\end{equation}
and the charge–voltage relation of a capacitor,
\begin{equation}
\label{eq:Capacitorcharge}
V_\varphi=\frac{1}{C_\varphi}\int_0^t i_\varphi(t')dt',
\end{equation}
which can be directly handled by standard SPICE-type circuit solvers. Equations \ref{eq:Josephson2int} and \ref{eq:Capacitorcharge} become formally equivalent under the identifications $i_\varphi(t)\equiv V(t)$ and $C_\varphi^{-1}\equiv 2\pi/\Phi_0$. Consequently, imposing on the solver the coupled solution of an additional loop (depicted in purple in \autoref{fig:fig1}d), in which an ideal current source charges a capacitor $C_\varphi$ with current $i_\varphi(t)$, the RSJ model can be fully implemented and solved. The combination of the phase-dynamics loop (purple) and the current–voltage dynamics loop (blue) yields a gate-electrode-tunable SNS Josephson junction, for which we introduce the circuit symbol shown in \autoref{fig:fig1}d.

A current-biased SNS junction typically exhibits hysteretic behavior: the switching current associated with the transition from the superconducting to the resistive state, namely, the critical current $I_c$, is systematically larger than the trapping current $I_r$, which corresponds to the transition from the normal to the superconducting state. This hysteresis arises from carrier heating in the dissipative regime. Incorporating such behavior into a circuit-level description, as employed here, is challenging because it would require a quantitatively accurate, time-resolved model of electron heating during the S-to-N switching event and subsequent electronic cooling dynamics. 
However, this level of microscopic modeling is not required. It is always possible to introduce an additional shunt resistor with resistance $R_S \ll R_N$ into the circuit. This strategy, which renders the junction overdamped \cite{BaronePaterno1982} and thereby removes the hysteresis, is commonly adopted in superconducting nanowire devices to maintain a low electronic temperature. It achieves this both by strongly reducing the duration over which the bias current dissipates Joule power on $R_N$ and by providing an effective cooling fin that enhances thermal coupling between electrons and substrate phonons \cite{Hao2024}. As discussed above, an SRO necessarily requires a shunted junction; thus, condition $I_r = I_c$ is automatically fulfilled if $R_S$ is chosen to be much smaller than $R_N$. Taking these considerations into account, the complete circuit representation of our device, shown in \autoref{fig:fig1}d, incorporates the appropriate shunt resistance and an ideal current-bias source.

Together, these ingredients allow one to calculate the time-dependent voltage-current characteristic $V(i(t))$ of the gate-controlled SNS. In what follows, we assume a specific material system that is archetypal and has been widely demonstrated experimentally. In particular, we assume that the superconducting electrodes are made of aluminum with a $T_c$ of 1.2 K. Moreover, we assume that the host layer of the 2DEG is a 100--nm-long and $1-\mu$m-wide InAs mesa, with carrier density $n_0=10^{12}$ cm$^{-2}$, electrically isolated from the gate electrode by a 50--nm-thick InAlAs layer characterized by $\epsilon_r=13$. These choices, compatible with an InAs quantum well in an In$_x$Al$_{1-x}$As heterostructure, yield the expected values $I_c\sim10$ $\mu$A, $R_N\sim10$ $\Omega$ and $C_g\simeq2.3$ fF. From these values, one can then estimate the expected Josephson energy $E_j=\frac{\Phi_0 I_c}{2\pi}\sim2\times10^{3} k_B T_c$, which confirms the suitability of the chosen system for executing devices with negligible thermal fluctuations and high phase coherence. Finally, the value of $R_N$ constrains the choice of $R_s$, which we set to 1 $\Omega$.

\begin{figure}[ht!]
\includegraphics[width=\linewidth]{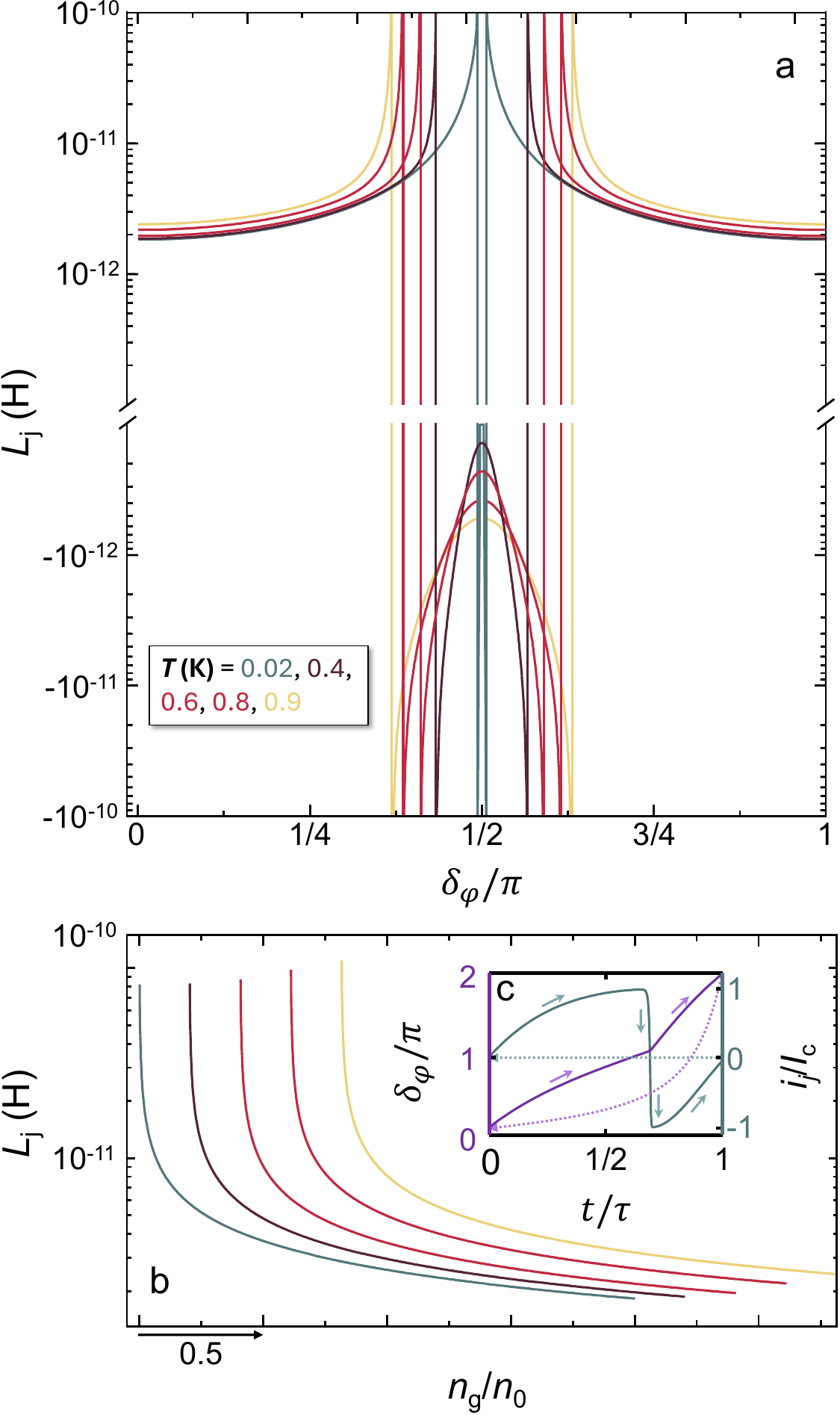}
\caption{\label{fig:Lj} 
Josephson inductance of a ballistic SNS junction and its impact on JRO dynamics.  
(a) Josephson inductance $L_j$ as a function of the superconducting phase difference $\delta_\varphi$ for selected temperatures, calculated from the ballistic KO-2 current–phase relation. The divergence of $L_j$ occurs at the inversion point of the Josephson current and shifts with temperature in accordance with the evolution of the current–phase relation (C$\Phi$R).  
(b) Josephson inductance as a function of the normalized gate-induced carrier density $n_g/n$ for selected temperatures. Due to the electrostatic tunability of the ballistic SNS junction, $L_j$ can be continuously tuned via the gate electrode and diverges as the two-dimensional electron gas approaches full depletion ($n_g/n \rightarrow -1$). For visual clarity, the curves are horizontally offset.  
(c) Temporal evolution of the superconducting phase difference $\delta_\varphi(t)$ and the Josephson current $i_j(t)$ over one oscillation period for $T = 20$ mK and $V_g = 0$, corresponding to the green curve in panel (a). The inversion of $i_j$ occurs near $\delta_\varphi \approx \pi$, where the Josephson inductance diverges.
}
\end{figure}

\autoref{fig:fig2} shows $V(t)$ of the shunted gate--controlled SNS computed for these characteristic parameter values as a function of a bias current $i(t)$ swept from 0 to 30 $\mu$A in 100 ns (panel a), for selected values of $V_g$ and $T$. To ensure the temporal resolution required for an accurate description of the device oscillations, such time-domain waveforms were obtained from transient simulations performed with a sub-picosecond time step. \autoref{fig:fig2}b shows $V(i(t))$ at 20 mK for gate voltages ranging from -0.7 to 0.6 V. The curves show a non-dissipative region ($V\approx0$, gray area in \autoref{fig:fig2}b) at low values of $i$. For currents exceeding $I_c(V_g)$, typical oscillations of a shunted Josephson junction appear, corresponding to the oscillatory partition of the current between the dissipative branches and the Josephson branch. Indeed, the onset of oscillations allows us to define the device's critical current operationally. As expected, the latter is controlled by the gate voltage, which, when increased from 0 V to 0.6 V, increases from its resting value of $\sim 12$ $\mu$A to 18 $\mu$A, due to charge accumulation in the 2DEG. Similarly, for negative gate voltages, the well is depleted, reducing the critical current, which is completely quenched at $\sim -0.7$ V.

\autoref{fig:fig2}c displays $V(i(t))$ traces for temperatures between 20 mK and 1.2 K at $V_g=0$. As expected, the critical current follows the temperature dependence predicted by the KO-2 model, indicating an essentially non-saturating behavior at low temperature  \cite{RevModPhys.76.411} and a complete suppression at the critical temperature $T_c \approx 1.2$ K. A comparison between datasets acquired at different gate voltages and those obtained at different temperatures is particularly illuminating. Both temperature and charge-carrier density govern not only the absolute value of the critical current but also the amplitude of the relaxation oscillations: these oscillations become more pronounced when the critical current is large and progressively diminish as it is reduced.
For gate-voltage-dependent curves, the oscillation waveform is essentially independent of $V_g$, exhibiting a nearly instantaneous rise as the junction switches to the normal state, followed by a slow decay as the current through the shunt resistance is redirected back into the junction. In contrast, in temperature-dependent characteristics, the waveform (i.e., the spectral content) of the oscillations evolves with increasing temperature, showing progressively smoother peaks as $T$ approaches $T_c$. This behavior can be attributed to the dual role of temperature in modifying the amplitude of the superconducting gap, which, in turn, affects both the magnitude of $I_c$ and C$\Phi$R. Although the first effect is analogous to the dependence of $I_c$ on the number of channels $N$ controlled by $V_g$, the latter has no equivalent in the action of the gate.

The observation of the characteristics $V$--$I(t)$ makes it clear that the period of the relaxation oscillations is essentially determined by the ratio between $i$ and $I_c$, which therefore represents the main knob through which the spectral characteristics of an SRO can be controlled. In general, $i$ and $T$ are the only quantities that can be controlled during device operation. Instead, the JRO can also exploit the additional control provided by the gate electrode. \autoref{fig:fig3} provides a quantitative comparison of the effects of these three degrees of freedom on the output voltage $V$. Similarly to what was observed for the $V$--$i(t)$, constant-current oscillations also exhibit relaxation oscillations, characterized by a rapid rise in the voltage across the junction, followed by a relatively slow relaxation phase, which is mainly responsible for the duration of the oscillation cycle. The data shown are obtained by selecting the final portion of a time-resolved simulation long enough to eliminate circuit transients; therefore, they represent the steady-state oscillations of the JRO. 
Since the duration of the transient depends on the biasing parameters, temperature, and gate settings of the device, to make the evolution of $V$ as a function of these parameters easier to interpret, the curves shown have been phase-shifted so that their first displayed peak coincides with time. Unless otherwise specified, this representation technique has also been used for all the data shown hereafter. 

Panel a shows a color plot of $V(i,t)$ for $T=20$ mK and $V_g=0$. The cut lines of $V(T)$ are shown for the selected currents in panel b. As $i$ increases, the period $\tau$ decreases as the ratio $i/I_c$ increases. In particular, $\tau$ can be continuously tuned by approximately a factor of 5 (from $\sim 1$ ns to a few hundred ps) over the bias-current range between $\sim I_c$ and $\ sim 2I_c$. The same effect is obtained in a completely analogous manner by applying a voltage to the gate electrode. \autoref{fig:fig3}c and d show the color plot and the selected cut lines of $V(V_g,t)$ for $i=20\mu$A and $T=20$ mK. 
In addition, in this case, the ratio $i/I_c$ controls the oscillation period through the action of the gate on the charge density of the 2DEG and on its critical current. $\tau$ decreases by approximately 65$\%$ as $V_g$ increases from -0.65 V (corresponding to an almost completely depleted 2DEG) to 0.65 V. Unlike the case of constant gate voltage, however, the oscillation amplitude is also significantly affected by the gate electrode, showing a reduction of approximately 40$\%$. Finally, in the temperature range of 20 mK to 1 K, for constant $i$ and $V_g$ fixed at 20$\mu$A and 0 V, respectively, $\tau$ shows (color plot and cut lines in \autoref{fig:fig3}, panels f to g, respectively) a rather limited reduction ($\sim15\%$) as $T$ increases, accompanied by a pronounced decrease in the oscillation amplitude ($\sim75\%$). However, as already discussed in relation to the characteristics $V$-- $i(t)$, increasing temperature affects the spectral characteristics of $V(t)$, which exhibits a “sinusoidalization” of the signal caused by a symmetrization of the rise and relaxation phases of the voltage across the JRO.

Since the modeled circuit is ideal, insofar as parasitic capacitances and inductances were not included in the calculation, its spectral response is entirely determined by the ratio between the Josephson inductance of the junction and the resistance of the dissipative branches of the device. It is worth highlighting that neither $L_J$ nor $R$ remains constant during the device operating cycle, and estimating the oscillation period using \textit{static} values of these quantities may be misleading and, in fact, underestimate the characteristic relaxation time of the device. Indeed, the oscillation of the Josephson current amplitude (and of the current flowing through the shunt resistance) corresponds to a cyclic opening and closing of the non-dissipative channel. Furthermore, the Josephson inductance, defined as
\begin{equation}
L_j=\frac{\hbar}{2e \frac{\partial\ i_j}{\partial\delta_\varphi}},
\end{equation}
and shown for different temperatures in \autoref{fig:Lj}a, can, in the case of a ballistic SNS junction, be analytically expressed as
\begin{equation}
\begin{aligned}
&L_j(\delta_\varphi) =\frac{I_c(T)}{2}\times\\
&\bigg[
\cos\left(\frac{\delta_\varphi}{2}\right)\tanh\left(\frac{\Delta(T)}{2 k_B T}\cos\left(\frac{\delta_\varphi}{2}\right)\right)+ \\
&-\frac{\Delta(T)}{2 k_B T}\sin^2\left(\frac{\delta_\varphi}{2}\right)
\left(1-\tanh^2\left(\frac{\Delta(T)}{2 k_B T}\cos\left(\frac{\delta_\varphi}{2}\right)\right)\right)
\bigg].
\end{aligned}
\end{equation}
$L_j$ is a function of the phase drop across the junction, which in turn evolves over time following the evolution of the Josephson current $i_j(t)$. The inset of \autoref{fig:Lj}c shows, as an example, the temporal evolution of $\delta_\varphi$ and $i_j$ over a period, calculated for $T=20$ mK and $V_g=0$, corresponding to the green curve in \autoref{fig:Lj}a. It can be observed that at the inversion point of $i_j$, which in this case occurs at $\delta_\varphi\sim\pi$, a divergence takes place in $L_j$, allowing it to explore an inductance range spanning 3 orders of magnitude or more (from $\sim 1$ pH up to 1 nH and above). The same occurs at all temperatures for which a Josephson current exists, with the divergence shifting to different points of the C$\Phi$R. The entire evolution of both inductance and resistance over one period collectively determines $\tau$, which, for a JRO sized as in our example, is around 1 ns. It is interesting to emphasize that in a JRO the gate voltage in turn controls, through $n$, the Josephson inductance, shown as a function of $n_g/n_0$ in \autoref{fig:Lj}c for selected temperatures, which, similarly to $R_N$, diverges for $n_g\rightarrow -n_0$.

\section{Voltage-controlled radiation comb generation}
\begin{figure}[t!]
\includegraphics[width=0.97\linewidth]{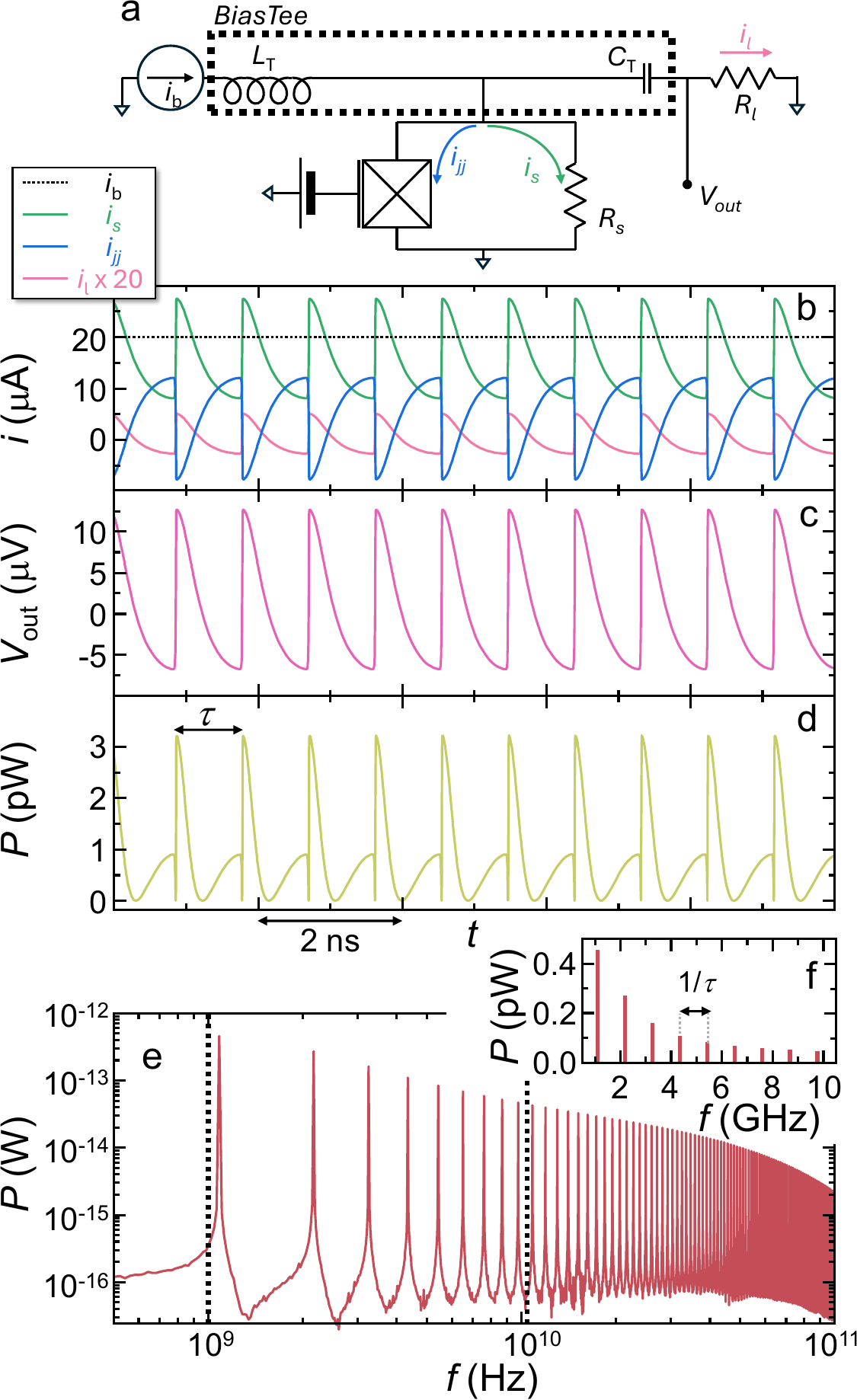}
\caption{\label{fig:fig4} 
Microwave frequency-comb generation in the Josephson relaxation oscillator (JRO). (a) Schematic representation of the JRO integrated with a bias–tee network comprising an inductance $L_T$ and a capacitance $C_T$, and terminated by a $50~\Omega$ load resistor $R_l$. (b) Steady-state temporal evolution of the currents flowing through the different branches of the circuit for $i=20~\mu$A (dashed black curve), $V_g=0$, and $T=20$ mK. The plot illustrates the periodic redistribution of current between the Josephson junction ($i_j$, blue curve), the shunt resistor ($i_s$, green curve), and the load ($i_l$, pink curve). (c) Output voltage $V_{\mathrm{out}}(t)$ measured across the load resistor. (d) Instantaneous power delivered to the load as a function of time $t$ for the same parameter values as in panels a and b. (e) Emission power spectrum obtained from the discrete fast Fourier transform of the output signal. (f) Linear-scale magnification of the low-frequency region of the spectrum corresponding to the interval delimited by the dashed grey lines in panel e. The comb lines are uniformly spaced by $\Delta f = 1/\tau$, where $\tau$ denotes the relaxation-oscillation period.
}
\end{figure}

Starting from the JRO, it is natural to consider implementing a comb generator, owing to the intrinsically periodic and self-sustained nature of the oscillation produced by such a circuit. Indeed, the nonlinear dynamics governing the JRO operation give rise to a periodic waveform in the time domain, which consequently translates into a spectrum rich in harmonics. This behavior makes the JRO an attractive candidate for comb-generation applications and motivates a deeper investigation of its spectral characteristics. In the following, the JRO discussed so far will therefore be numerically characterized from this perspective, with particular attention devoted to its capability to generate a harmonic comb at the output.

To properly extract and analyze the AC signal generated by the oscillator, a bias--tee network is introduced into the circuit (see \autoref{fig:fig4}a). The values adopted for the bias--tee network, namely $L_T=100\,\mathrm{nH}$ and $C_T=30\,\mathrm{nF}$, were chosen to ensure a clear separation between the DC bias path and the AC oscillating component over the frequency range of interest for quantum technology and computing, which we identify in the $\lesssim 10$ GHz range. Although the purpose of the present study is not the detailed optimization of the bias network itself, but rather the numerical characterization of the comb-generation capability of the JRO, the introduction of the bias--tee also enables the model to naturally account for parasitic inductive contributions that are unavoidably present in realistic implementations, such as bond--wire inductances, low-pass filtering elements along the bias line, and other stray series inductances. Nevertheless, it is worth noting that in a fully realistic microwave implementation extending up to $10\,\mathrm{GHz}$ or higher, lumped elements with such values may exhibit non-idealities due to parasitic capacitances and self-resonant effects. Finally, as a representative output termination, a standard 50 $\Omega$ load resistor $R_l$ is used, consistent with typical microwave and RF measurement environments. 

\autoref{fig:fig4}a shows the oscillatory behavior of the JRO in its operating configuration, \textit{i. e.}, integrated into the circuit equipped with a bias tie and load resistance, for a bias current of 20~$\mu$A, $V_g = 0$~V and $T = 20$~mK. The plot shows how the current is distributed over time between the Josephson branches (blue curve) and the dissipative branches of the device. The current peak across $R_l$ ($i_l$, pink curve) is approximately 20 times smaller than that across $R_s$ ($i_s$, green curve). This is due to the sizes of $R_s$ and $R_l$ and represents an intrinsic limitation of the chosen configuration, since a significant portion of the energy dissipated per operating cycle is wasted in the shunt resistance rather than delivered to the load. This limitation comes directly from the value of $R_N$, which in turn imposes an upper bound on $R_s$. However, by designing the junction so that its normal-state resistance is naturally much larger than that of the chosen load, it is, in principle, possible to completely eliminate the shunt resistance and make the entire power generated by the JRO available for use. Nevertheless, we chose to retain the configuration presented here because it still provides a lower bound on the output power and preserves a universal setup, given the conventional value of the load resistance.

The oscillations of the output voltage, \textit{that is,} across $R_l$, are shown in \autoref{fig:fig4}c for the same values of $i$, $V_g$, and $T$ as in panel b. Thanks to the high-pass stage of the bias tee, the oscillation has essentially zero time average, is in phase with $i_l$, and has a peak-to-peak amplitude of $\sim 30\,\mu\mathrm{V}$. The product $V_{\mathrm{out}}(t)\, i_l(t)$ gives the instantaneous power $P(t)$ delivered to the load and is shown in \autoref{fig:fig4}d. For the parameter values chosen in this example, the generated power reaches a peak of 3~pW, with a total energy dissipated over a period $\tau$ ($\simeq1$ ns) of $\sim 10^{-21}$~J.

The corresponding emission spectrum of the signal is shown in \autoref{fig:fig4}e. It was obtained by applying a discrete Fast Fourier Transform (FFT) to $P(t)$ according to
\begin{equation}
\tilde{P}(f_k)=\sum_{j=0}^{N-1} P(t_j)\,
e^{-i2\pi kj/M},
\end{equation}
where $M$ is the total number of uniformly sampled temporal points and $f_k = k/(M\Delta t)$ is the discrete frequency axis associated with the sampling interval $\Delta t$. The transient response was removed from the temporal traces by selecting only the portion of the waveform between the first and the last detected local maxima of $V_{\mathrm{out}}(t)$ in steady-state oscillation. This ensured that the time domain spanned an integer multiple of $\tau$, thereby eliminating spectral leakage and making an additional windowing function unnecessary. The spectral amplitude was then computed as
\begin{equation}
P(f_k)=\frac{1}{N}\left|\tilde{P}(f_k)\right|.
\end{equation}
Only the positive-frequency components of the FFT were retained in the final spectra. The resulting quantity $P(f)$ therefore represents the relative spectral distribution of the emitted power, which can be used to identify the oscillation modes and their evolution as a function of the biasing parameters. As anticipated, the power spectrum of the JRO implements a frequency comb whose peaks extend from $\sim 1$~GHz upward, with peak power per mode decreasing from about half a pW at the fundamental to roughly one-tenth of a pW at 10~GHz, remaining above 10~fW at 100~GHz. As expected, the oscillation modes are equally spaced in frequency by $\Delta f = \frac{1}{\tau}$, which in the example of \autoref{fig:fig4} corresponds to $\sim 1$~GHz, as can be seen in \autoref{fig:fig4}f, which shows a linear-scale zoom of the spectrum between 100~MHz and 10~GHz.
\begin{figure}[t!]
\includegraphics[width=\linewidth]{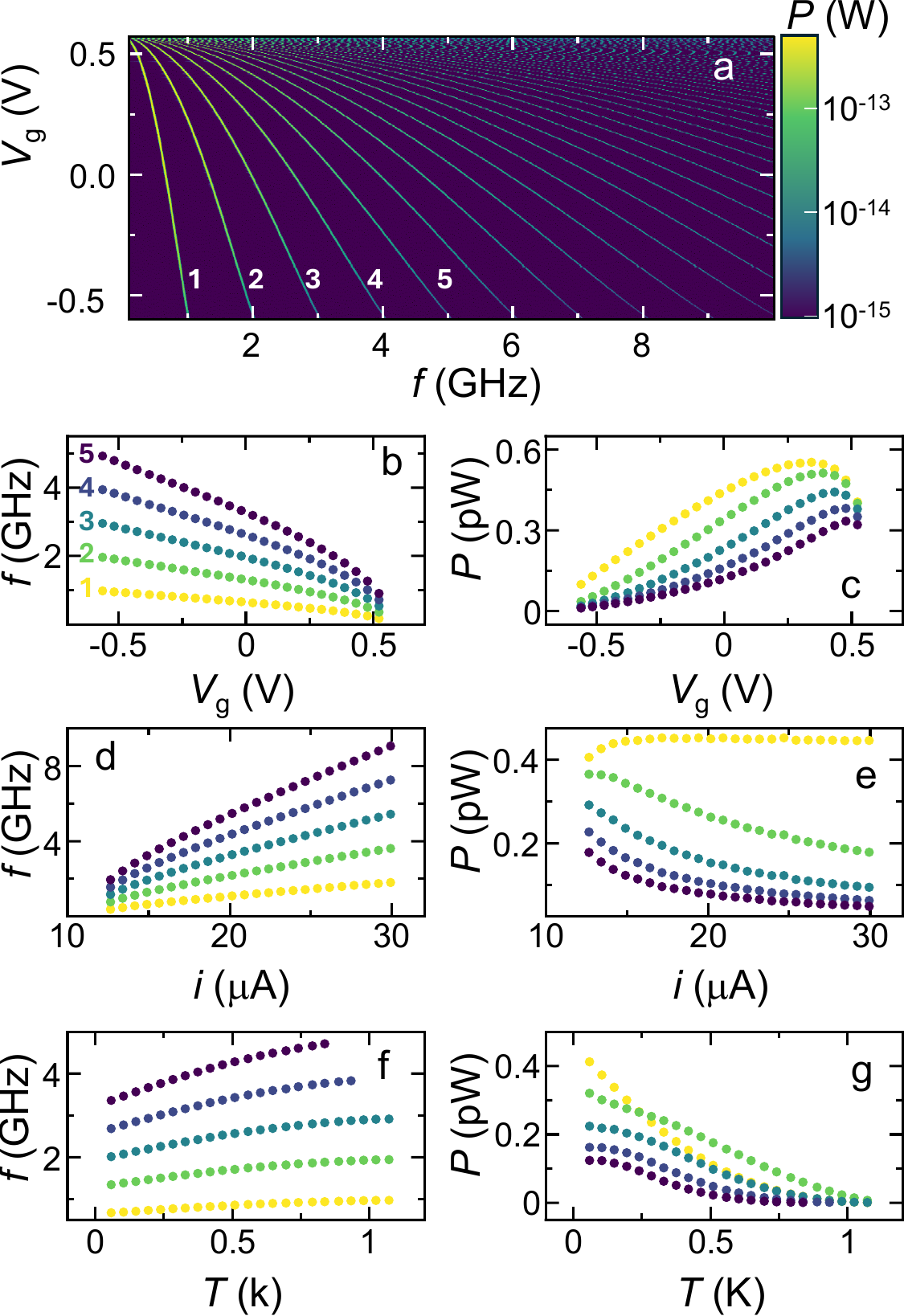}
\caption{\label{fig:fig5} 
Gate-, current-, and temperature-dependent control of the microwave frequency comb generated by the JRO. (a) Color map of the emitted power spectrum in the range 1–10 GHz as a function of gate voltage $V_g$, computed for a bias current $i = 15~\mu\text{A}$ and temperature $T = 20$ mK. (b) Frequencies $f_m$ of the first five comb modes ($m = 1,\ldots,5$), extracted from panel (a), plotted as a function of gate voltage. (c) Corresponding peak powers $P_m$ of the first five comb modes as a function of gate voltage. (d) Frequencies $f_m$ of the first five comb modes as a function of bias current for $V_g = 0$ and $T = 20$ mK. (e) Peak powers $P_m$ of the first five comb modes as a function of bias current. (f) Frequencies $f_m$ of the first five comb modes as a function of temperature for $V_g = 0$ and $i = 15~\mu\text{A}$. (g) Peak powers $P_m$ as a function of temperature.
}
\end{figure}

At this stage, it is instructive to examine how the JRO emission spectrum depends on its three main control parameters, namely the bias current $i$, the temperature $T$, and, most importantly, the gate voltage $V_g$. Figure~\autoref{fig:fig5}a displays a color map of the emission spectrum in the 1--10~GHz frequency range as a function of $V_g$, measured at $i=15\,\mu\mathrm{A}$ and $T=20\,\mathrm{mK}$. The mode spacing increases as the gate voltage is reduced, evolving from an almost complete collapse of the oscillation modes for $V_g \gtrsim 0.5\,\mathrm{V}$, where the increase of $I_c(V_g)$ effectively suppresses the oscillatory regime by exceeding the applied bias current $i$, to a mode spacing of approximately 1~GHz near the critical--current pinch-off, at $V_g \lesssim -0.5\,\mathrm{V}$. Remarkably, this behavior implies that, by appropriately tuning $V_g$, it is always possible to identify at least one emission mode at any desired frequency within the overall emission bandwidth of the JRO. In particular, continuous frequency coverage can be achieved throughout the technologically relevant 1--10~GHz band.   

To enable a more quantitative analysis of the evolution of the spectrum, we focus on the first 5 modes, labeled by their index $m={1,..,5}$ in \autoref{fig:fig5}a, and we plot their frequency $f_m$ and peak power $P_m$ as a function of gate voltage, bias current, and temperature in \autoref{fig:fig5}b and c, d and e, and f and g, respectively. As already observed, $f_m$ and the frequency spacing between consecutive modes ($f_{m+1}-f_{m}$) increase as $V_g$ decreases, ensuring that at least one emission line is always available within the 1-5 GHz range. However, from the point of view of emission power, the behavior as a function of gate voltage is more peculiar, showing a non-monotonic evolution for all 5 modes with a peak around $\sim0.4$ V. This behavior is due to the interplay of two factors: the emission power increases with increasing charge density in the semiconductor channel, but when $I_c$ approaches and eventually exceeds $i$, the oscillation must damp and then turn off.

The evolution of $f_m$ with bias current is analogous to that obtained with the gate voltage. However, the effect on the emission power of the individual modes is quite different. Indeed, a clear damping of the modes with index $\geq2$ is present, while $P_1(i)$ appears to saturate substantially in the current bias range explored for $i\gtrsim15$ A.

Finally, a more limited variation is observed in the frequency of the individual modes and in their spacing \textit{vs.} $T$, compared to the case of varying $V_g$ and $i$. However, increasing $T$ from 20 mK to $T_c$ reveals a marked redistribution of power among modes, with even a crossover between mode 1 and modes 2 and 3. This provides quantitative confirmation of how strongly the effect of temperature on C$\Phi$R directly reflects onto the spectrum of relaxation oscillations.

\section{Niobium-based JROs}
\begin{figure}[t!]
\includegraphics[width=\linewidth]{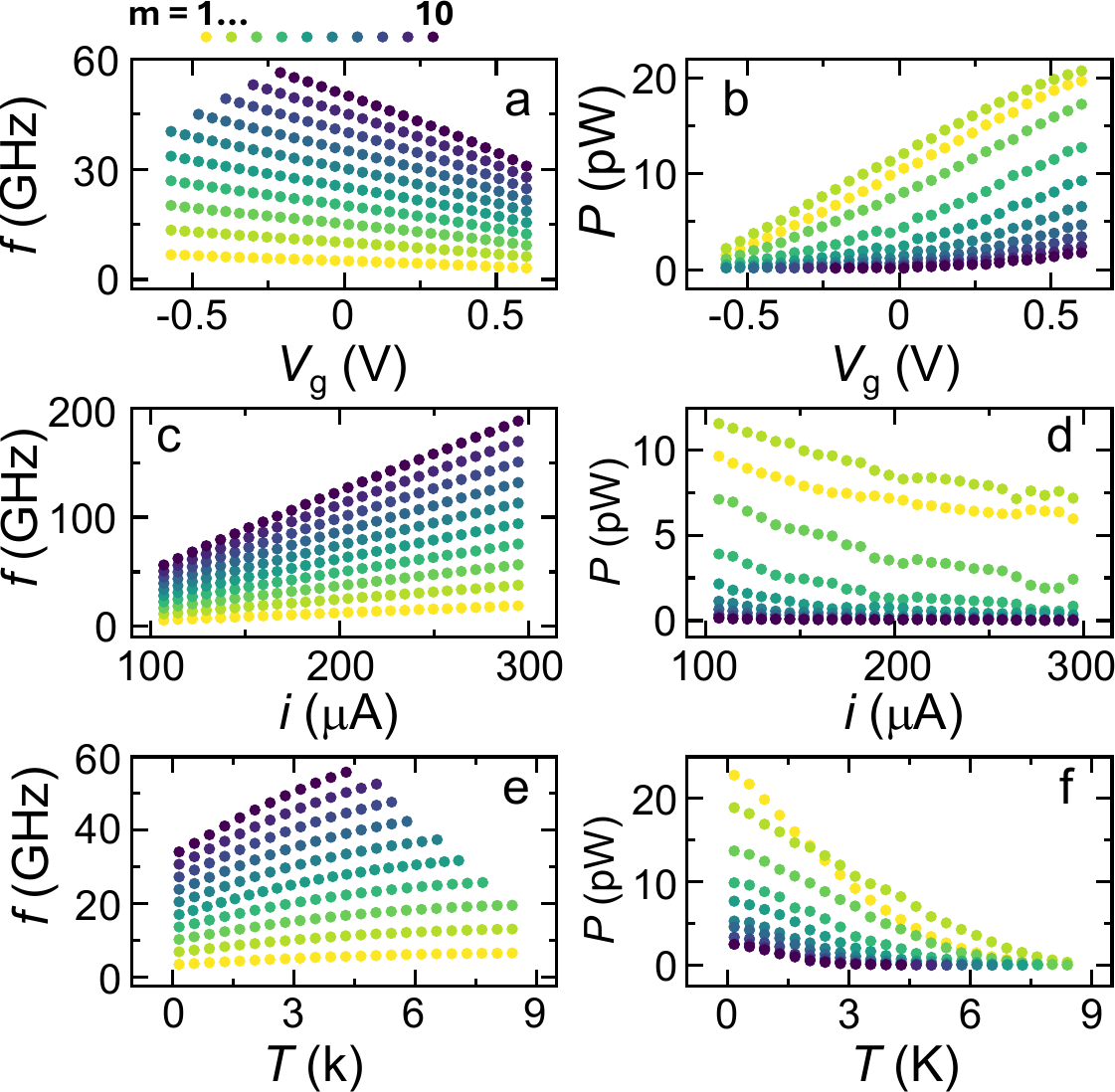}
\caption{\label{fig:figNb} 
Performance characteristics of a niobium-based Josephson radiation oscillator (JRO). The device parameters are taken to be identical to those of the Al/InAs implementation, except that the superconducting electrodes are modeled as elemental Nb with a critical temperature $T_c = 9.2$ K. (a) Frequencies $f_j$ of the first five comb modes ($m = 1,\ldots,10$) as a function of gate voltage for $T = 3$ K and bias current $i = 100~\mu$A. (b) Corresponding peak powers $P_j$ of the first ten comb modes as a function of gate voltage. (c) Frequencies $f_j$ of the first ten comb modes as a function of bias current for $V_g = 0$ and $T = 3$ K. (d) Peak powers $P_j$ of the first ten comb modes as a function of bias current. (f) Frequencies $f_j$ of the first ten comb modes as a function of temperature for $V_g = 0$ and $i = 100~\mu$A. (g) Peak powers $P_j$ as a function of temperature.
}
\end{figure}
Although the present analysis has been carried out for an Al/InAs Josephson field-effect transistor, the Josephson Radiation Oscillator (JRO) concept is not intrinsically restricted to aluminum-based weak links. An especially compelling generalization is the implementation of an analogous device architecture employing niobium (Nb) superconducting electrodes. Because of its larger superconducting energy gap and higher critical temperature, Nb is expected to sustain significantly higher critical currents and Josephson energies, thereby enabling operation at elevated temperatures while maintaining robust phase coherence. As a result, Nb-based JROs are expected to generate microwave-frequency combs with increased output power, extended accessible frequency ranges, and improved resilience against thermal fluctuations.

From a technological standpoint, the extension of the JRO platform to niobium is particularly advantageous because it relaxes the stringent cryogenic requirements commonly associated with aluminum-based implementations. Although Al-based realizations are intrinsically optimized for operation in millikelvin environments and within quantum-computing infrastructures, Nb-based architectures are expected to function efficiently in the few-kelvin regime routinely achievable with compact, closed-cycle cryocoolers. This relaxed cooling requirement substantially broadens the range of feasible applications, encompassing cryogenic microwave instrumentation, superconducting digital electronics, frequency-multiplexed sensor readout architectures, and neuromorphic superconducting circuitry. In addition, the higher characteristic frequencies afforded by niobium are anticipated to significantly expand the accessible comb bandwidth and to enable seamless integration with established Nb-based superconducting technologies, including single-flux-quantum logic, Josephson voltage standards, and superconducting microwave integrated circuits.

Figures~\autoref{fig:figNb}(a), (c) and (e) report the frequencies $f_j$ of the first ten comb modes, labeled by the mode index $m=\{1,\ldots,10\}$, as a function of gate voltage $V_g$, normalized bias current $i$, and temperature $T$, respectively, for a device identical to that considered above except for the use of elemental Nb electrodes with critical temperature $T_c=9.2$~K. All remaining device parameters are left unchanged. The corresponding power $P_j$ dissipated in the load resistor at the central frequency of each mode is shown in panels (b), (d), and (f) as a function of $V_g$, $i$, and $T$, respectively.

The overall behavior of the Nb-based JRO closely resembles that observed for the Al counterpart. The most notable difference is the substantially larger frequency span, which extends from approximately $6$~GHz up to nearly $100$~GHz for the first five modes. Within this range, the powers emitted are roughly comprised between $1$ and $10$~pW. For characterization as a function of $V_g$ and $i$, the temperature has been fixed at $T=3$~K, a particularly relevant value since it can be routinely achieved in closed-cycle cryogenic systems $^4$He without significant experimental complexity.

As a consequence of this higher operating temperature, the spectral evolution as a function of the gate voltage does not exhibit exactly the same features observed in the aluminum case. Nevertheless, the main tuning mechanism remains unchanged: both the mode frequencies $f_j$ and the spacing between adjacent modes, $f_{j+1}-f_j$, increase as $V_g$ decreases, as shown in Figure~\autoref{fig:figNb}(a). In contrast to the Al-based device, the power emitted in Figure~\autoref{fig:figNb}(b) no longer exhibits a distinct maximum as a function of the gate voltage. Instead, it increases monotonically with increasing carrier density and gate voltage. In particular, the second mode (light-green markers) becomes more powerful than the fundamental one. This behavior originates from the crossing between the first and second modes previously observed in the aluminum device around $T\simeq T_c/3$ (see also Figure~\autoref{fig:figNb}(f)). Remarkably, the emitted power is approximately 30 times larger than that obtained in the corresponding Al implementation.

As illustrated in Figure~\autoref{fig:figNb}(c), a further expansion of the frequency range can be achieved by increasing the bias current at fixed $V_g=0$. However, this enhancement comes at the expense of the emitted power. Indeed, the power associated with all the modes investigated, as shown in Figure~\autoref{fig:figNb}(d), decreases slightly but systematically as the bias current increases. A similar trend emerges as the temperature increases. The resulting reduction in the ratio $I_c/i$ shifts the comb modes toward higher frequencies, as reported in Figure~\autoref{fig:figNb}(e). However, the emitted power is significantly suppressed above approximately $T_c/2$, as shown in Figure~\autoref{fig:figNb}(f). Beyond this threshold, only the first five modes remain appreciably populated.

Overall, these results indicate that the Nb-based JRO preserves the fundamental tuning mechanisms previously established in analogous aluminum-based devices, while simultaneously extending the accessible frequency bandwidth and enhancing robustness against temperature-induced variations. These performance characteristics are a direct consequence of the larger superconducting energy gap and higher critical temperature of niobium, underscoring its suitability as a resilient platform for coherent Josephson radiation sources that operate at higher frequencies and elevated temperatures.

\section{Conclusions}

In conclusion, we have introduced and numerically characterized a gate-tunable Josephson Radiation Oscillator (JRO) based on a ballistic superconducting–normal–superconducting (SNS) Josephson field-effect transistor operated in the relaxation-oscillation regime. By integrating the nonlinear dynamics of a resistively shunted Josephson junction with the electrostatic tunability of a semiconducting weak link, this device architecture realizes a phase-coherent microwave frequency comb. The spectral properties of the generated comb are continuously adjustable via a gate electrode.

Time-domain SPICE simulations demonstrate that the Josephson relaxation oscillator (JRO) exhibits stable relaxation oscillations over a technologically relevant gigahertz (GHz) frequency range. The resulting frequency comb consists of equidistant spectral lines whose central frequencies, spacing, and emitted power can be dynamically tuned in real time via three independent control parameters: gate voltage, bias current, and temperature. Among these, the gate voltage is the principal control knob of the system, enabling direct electrical modulation of the critical current, Josephson inductance, oscillation period, and the overall comb spectrum, without requiring magnetic fields, cavity engineering, or thermal tuning. For a representative device implementation based on aluminum (Al) and indium arsenide (InAs), this mechanism enables continuous spectral coverage from 1 to 10 GHz, such that at least one comb line can be placed at any target frequency within this interval.
The analysis highlights the pivotal role of the strongly nonlinear Josephson inductance in ballistic superconductor–normal metal–superconductor (SNS) junctions. Its evolution over a full oscillation cycle is central to establishing the characteristic relaxation dynamics that generate the frequency comb. In addition, the temperature dependence of the current–phase relation introduces an extra degree of freedom for tailoring both the modal power distribution and the spectral composition of the emitted radiation.

We have demonstrated that the underlying concept is readily transferable to other superconducting material platforms. In particular, niobium-based implementations preserve the same tuning mechanisms while enabling a substantial extension of both the accessible frequency range and the radiated power. Owing to niobium’s larger superconducting energy gap and elevated critical temperature, the comb bandwidth can be increased to approximately 100 GHz while maintaining operation at temperatures of only a few kelvins. This development markedly relaxes the associated cryogenic constraints.

The proposed Josephson relaxation oscillator (JRO) integrates the structural compactness characteristic of a single mesoscopic Josephson device with the enhanced tunability afforded by electrostatic gating. This combination yields a scalable platform for generating tunable microwave frequency combs. These properties render the JRO a promising candidate for implementation in superconducting quantum information architectures, cryogenic microwave electronics, multiplexed sensor readout schemes, single-flux-quantum (SFQ) logic systems, and neuromorphic superconducting circuit technologies. Collectively, the results establish that Josephson relaxation oscillations in gate-tunable weak links provide a novel route toward electrically programmable, broadband, and fully integrated superconducting frequency-comb sources.

\acknowledgments

The authors acknowledge A. Crippa for useful discussions.


\begin{thebibliography}{72}%
\makeatletter
\providecommand \@ifxundefined [1]{%
 \@ifx{#1\undefined}
}%
\providecommand \@ifnum [1]{%
 \ifnum #1\expandafter \@firstoftwo
 \else \expandafter \@secondoftwo
 \fi
}%
\providecommand \@ifx [1]{%
 \ifx #1\expandafter \@firstoftwo
 \else \expandafter \@secondoftwo
 \fi
}%
\providecommand \natexlab [1]{#1}%
\providecommand \enquote  [1]{``#1''}%
\providecommand \bibnamefont  [1]{#1}%
\providecommand \bibfnamefont [1]{#1}%
\providecommand \citenamefont [1]{#1}%
\providecommand \href@noop [0]{\@secondoftwo}%
\providecommand \href [0]{\begingroup \@sanitize@url \@href}%
\providecommand \@href[1]{\@@startlink{#1}\@@href}%
\providecommand \@@href[1]{\endgroup#1\@@endlink}%
\providecommand \@sanitize@url [0]{\catcode `\\12\catcode `\$12\catcode `\&12\catcode `\#12\catcode `\^12\catcode `\_12\catcode `\%12\relax}%
\providecommand \@@startlink[1]{}%
\providecommand \@@endlink[0]{}%
\providecommand \url  [0]{\begingroup\@sanitize@url \@url }%
\providecommand \@url [1]{\endgroup\@href {#1}{\urlprefix }}%
\providecommand \urlprefix  [0]{URL }%
\providecommand \Eprint [0]{\href }%
\providecommand \doibase [0]{https://doi.org/}%
\providecommand \selectlanguage [0]{\@gobble}%
\providecommand \bibinfo  [0]{\@secondoftwo}%
\providecommand \bibfield  [0]{\@secondoftwo}%
\providecommand \translation [1]{[#1]}%
\providecommand \BibitemOpen [0]{}%
\providecommand \bibitemStop [0]{}%
\providecommand \bibitemNoStop [0]{.\EOS\space}%
\providecommand \EOS [0]{\spacefactor3000\relax}%
\providecommand \BibitemShut  [1]{\csname bibitem#1\endcsname}%
\let\auto@bib@innerbib\@empty
\bibitem [{\citenamefont {Udem}\ \emph {et~al.}(2002)\citenamefont {Udem}, \citenamefont {Holzwarth},\ and\ \citenamefont {H{\"a}nsch}}]{udem2002optical}%
  \BibitemOpen
  \bibfield  {author} {\bibinfo {author} {\bibfnamefont {T.}~\bibnamefont {Udem}}, \bibinfo {author} {\bibfnamefont {R.}~\bibnamefont {Holzwarth}},\ and\ \bibinfo {author} {\bibfnamefont {T.~W.}\ \bibnamefont {H{\"a}nsch}},\ }\href@noop {} {\bibfield  {journal} {\bibinfo  {journal} {Nature}\ }\textbf {\bibinfo {volume} {416}},\ \bibinfo {pages} {233} (\bibinfo {year} {2002})}\BibitemShut {NoStop}%
\bibitem [{\citenamefont {Burghoff}\ \emph {et~al.}(2015)\citenamefont {Burghoff}, \citenamefont {Yang}, \citenamefont {Hayton}, \citenamefont {Gao}, \citenamefont {Reno},\ and\ \citenamefont {Hu}}]{burghoff2015}%
  \BibitemOpen
  \bibfield  {author} {\bibinfo {author} {\bibfnamefont {D.}~\bibnamefont {Burghoff}}, \bibinfo {author} {\bibfnamefont {Y.}~\bibnamefont {Yang}}, \bibinfo {author} {\bibfnamefont {D.~J.}\ \bibnamefont {Hayton}}, \bibinfo {author} {\bibfnamefont {J.-R.}\ \bibnamefont {Gao}}, \bibinfo {author} {\bibfnamefont {J.~L.}\ \bibnamefont {Reno}},\ and\ \bibinfo {author} {\bibfnamefont {Q.}~\bibnamefont {Hu}},\ }\href {https://doi.org/10.1364/OE.23.001190} {\bibfield  {journal} {\bibinfo  {journal} {Opt. Express}\ }\textbf {\bibinfo {volume} {23}},\ \bibinfo {pages} {1190} (\bibinfo {year} {2015})}\BibitemShut {NoStop}%
\bibitem [{\citenamefont {Hillbrand}\ \emph {et~al.}(2020)\citenamefont {Hillbrand}, \citenamefont {Auth}, \citenamefont {Piccardo}, \citenamefont {Opa\ifmmode~\check{c}\else \v{c}\fi{}ak}, \citenamefont {Gornik}, \citenamefont {Strasser}, \citenamefont {Capasso}, \citenamefont {Breuer},\ and\ \citenamefont {Schwarz}}]{hillbrand2020prl}%
  \BibitemOpen
  \bibfield  {author} {\bibinfo {author} {\bibfnamefont {J.}~\bibnamefont {Hillbrand}}, \bibinfo {author} {\bibfnamefont {D.}~\bibnamefont {Auth}}, \bibinfo {author} {\bibfnamefont {M.}~\bibnamefont {Piccardo}}, \bibinfo {author} {\bibfnamefont {N.}~\bibnamefont {Opa\ifmmode~\check{c}\else \v{c}\fi{}ak}}, \bibinfo {author} {\bibfnamefont {E.}~\bibnamefont {Gornik}}, \bibinfo {author} {\bibfnamefont {G.}~\bibnamefont {Strasser}}, \bibinfo {author} {\bibfnamefont {F.}~\bibnamefont {Capasso}}, \bibinfo {author} {\bibfnamefont {S.}~\bibnamefont {Breuer}},\ and\ \bibinfo {author} {\bibfnamefont {B.}~\bibnamefont {Schwarz}},\ }\href {https://doi.org/10.1103/PhysRevLett.124.023901} {\bibfield  {journal} {\bibinfo  {journal} {Phys. Rev. Lett.}\ }\textbf {\bibinfo {volume} {124}},\ \bibinfo {pages} {023901} (\bibinfo {year} {2020})}\BibitemShut {NoStop}%
\bibitem [{\citenamefont {H{\"a}nsch}\ and\ \citenamefont {Walther}(1999)}]{hansch1999laser}%
  \BibitemOpen
  \bibfield  {author} {\bibinfo {author} {\bibfnamefont {T.}~\bibnamefont {H{\"a}nsch}}\ and\ \bibinfo {author} {\bibfnamefont {H.}~\bibnamefont {Walther}},\ }\href@noop {} {\bibfield  {journal} {\bibinfo  {journal} {Rev. Mod. Phys.}\ }\textbf {\bibinfo {volume} {71}},\ \bibinfo {pages} {S242} (\bibinfo {year} {1999})}\BibitemShut {NoStop}%
\bibitem [{\citenamefont {Bloembergen}(1982)}]{RevModPhys.54.685}%
  \BibitemOpen
  \bibfield  {author} {\bibinfo {author} {\bibfnamefont {N.}~\bibnamefont {Bloembergen}},\ }\href {https://doi.org/10.1103/RevModPhys.54.685} {\bibfield  {journal} {\bibinfo  {journal} {Rev. Mod. Phys.}\ }\textbf {\bibinfo {volume} {54}},\ \bibinfo {pages} {685} (\bibinfo {year} {1982})}\BibitemShut {NoStop}%
\bibitem [{\citenamefont {H{\"a}nsch}\ and\ \citenamefont {Inguscio}(1994)}]{hansch1994frontiers}%
  \BibitemOpen
  \bibfield  {author} {\bibinfo {author} {\bibfnamefont {T.~W.}\ \bibnamefont {H{\"a}nsch}}\ and\ \bibinfo {author} {\bibfnamefont {M.}~\bibnamefont {Inguscio}},\ }\href@noop {} {\emph {\bibinfo {title} {Frontiers in Laser Spectroscopy}}},\ Vol.\ \bibinfo {volume} {120}\ (\bibinfo  {publisher} {North Holland},\ \bibinfo {year} {1994})\BibitemShut {NoStop}%
\bibitem [{\citenamefont {Foreman}\ \emph {et~al.}(2007)\citenamefont {Foreman}, \citenamefont {Holman}, \citenamefont {Hudson}, \citenamefont {Jones},\ and\ \citenamefont {Ye}}]{foreman2007remote}%
  \BibitemOpen
  \bibfield  {author} {\bibinfo {author} {\bibfnamefont {S.~M.}\ \bibnamefont {Foreman}}, \bibinfo {author} {\bibfnamefont {K.~W.}\ \bibnamefont {Holman}}, \bibinfo {author} {\bibfnamefont {D.~D.}\ \bibnamefont {Hudson}}, \bibinfo {author} {\bibfnamefont {D.~J.}\ \bibnamefont {Jones}},\ and\ \bibinfo {author} {\bibfnamefont {J.}~\bibnamefont {Ye}},\ }\href@noop {} {\bibfield  {journal} {\bibinfo  {journal} {Rev. Sci. Instrum.}\ }\textbf {\bibinfo {volume} {78}},\ \bibinfo {pages} {021101} (\bibinfo {year} {2007})}\BibitemShut {NoStop}%
\bibitem [{\citenamefont {Picqu{\'e}}\ and\ \citenamefont {H{\"a}nsch}(2019)}]{reviewcomb_optics}%
  \BibitemOpen
  \bibfield  {author} {\bibinfo {author} {\bibfnamefont {N.}~\bibnamefont {Picqu{\'e}}}\ and\ \bibinfo {author} {\bibfnamefont {T.~W.}\ \bibnamefont {H{\"a}nsch}},\ }\href@noop {} {\bibfield  {journal} {\bibinfo  {journal} {Nature Photonics}\ }\textbf {\bibinfo {volume} {13}},\ \bibinfo {pages} {146} (\bibinfo {year} {2019})}\BibitemShut {NoStop}%
\bibitem [{\citenamefont {Barends}\ \emph {et~al.}(2013)\citenamefont {Barends}, \citenamefont {Kelly}, \citenamefont {Megrant}, \citenamefont {Sank}, \citenamefont {Jeffrey}, \citenamefont {Chen}, \citenamefont {Yin}, \citenamefont {Chiaro}, \citenamefont {Mutus}, \citenamefont {Neill}, \citenamefont {O'Malley}, \citenamefont {Roushan}, \citenamefont {Wenner}, \citenamefont {White}, \citenamefont {Cleland},\ and\ \citenamefont {Martinis}}]{martinis_Xmon}%
  \BibitemOpen
  \bibfield  {author} {\bibinfo {author} {\bibfnamefont {R.}~\bibnamefont {Barends}}, \bibinfo {author} {\bibfnamefont {J.}~\bibnamefont {Kelly}}, \bibinfo {author} {\bibfnamefont {A.}~\bibnamefont {Megrant}}, \bibinfo {author} {\bibfnamefont {D.}~\bibnamefont {Sank}}, \bibinfo {author} {\bibfnamefont {E.}~\bibnamefont {Jeffrey}}, \bibinfo {author} {\bibfnamefont {Y.}~\bibnamefont {Chen}}, \bibinfo {author} {\bibfnamefont {Y.}~\bibnamefont {Yin}}, \bibinfo {author} {\bibfnamefont {B.}~\bibnamefont {Chiaro}}, \bibinfo {author} {\bibfnamefont {J.}~\bibnamefont {Mutus}}, \bibinfo {author} {\bibfnamefont {C.}~\bibnamefont {Neill}}, \bibinfo {author} {\bibfnamefont {P.}~\bibnamefont {O'Malley}}, \bibinfo {author} {\bibfnamefont {P.}~\bibnamefont {Roushan}}, \bibinfo {author} {\bibfnamefont {J.}~\bibnamefont {Wenner}}, \bibinfo {author} {\bibfnamefont {T.~C.}\ \bibnamefont {White}}, \bibinfo {author} {\bibfnamefont {A.~N.}\ \bibnamefont {Cleland}},\ and\ \bibinfo {author} {\bibfnamefont {J.~M.}\ \bibnamefont
  {Martinis}},\ }\href {https://doi.org/10.1103/PhysRevLett.111.080502} {\bibfield  {journal} {\bibinfo  {journal} {Phys. Rev. Lett.}\ }\textbf {\bibinfo {volume} {111}},\ \bibinfo {pages} {080502} (\bibinfo {year} {2013})}\BibitemShut {NoStop}%
\bibitem [{\citenamefont {Crippa}\ \emph {et~al.}(2019)\citenamefont {Crippa} \emph {et~al.}}]{crippa2019gate}%
  \BibitemOpen
  \bibfield  {author} {\bibinfo {author} {\bibfnamefont {A.}~\bibnamefont {Crippa}} \emph {et~al.},\ }\href@noop {} {\bibfield  {journal} {\bibinfo  {journal} {Nature Communications}\ }\textbf {\bibinfo {volume} {10}},\ \bibinfo {pages} {2776} (\bibinfo {year} {2019})}\BibitemShut {NoStop}%
\bibitem [{\citenamefont {Koshelets}\ and\ \citenamefont {Shitov}(2000)}]{koshelets_integrated_2000}%
  \BibitemOpen
  \bibfield  {author} {\bibinfo {author} {\bibfnamefont {V.~P.}\ \bibnamefont {Koshelets}}\ and\ \bibinfo {author} {\bibfnamefont {S.~V.}\ \bibnamefont {Shitov}},\ }\href {https://doi.org/10.1088/0953-2048/13/5/201} {\bibfield  {journal} {\bibinfo  {journal} {Supercond. Sci. Technol.}\ }\textbf {\bibinfo {volume} {13}},\ \bibinfo {pages} {R53} (\bibinfo {year} {2000})}\BibitemShut {NoStop}%
\bibitem [{\citenamefont {Galin}\ \emph {et~al.}(2015)\citenamefont {Galin}, \citenamefont {Klushin}, \citenamefont {Kurin}, \citenamefont {Seliverstov}, \citenamefont {Finkel}, \citenamefont {Goltsman}, \citenamefont {Müller}, \citenamefont {Scheller},\ and\ \citenamefont {Semenov}}]{galin_towards_2015}%
  \BibitemOpen
  \bibfield  {author} {\bibinfo {author} {\bibfnamefont {M.~A.}\ \bibnamefont {Galin}}, \bibinfo {author} {\bibfnamefont {A.~M.}\ \bibnamefont {Klushin}}, \bibinfo {author} {\bibfnamefont {V.~V.}\ \bibnamefont {Kurin}}, \bibinfo {author} {\bibfnamefont {S.~V.}\ \bibnamefont {Seliverstov}}, \bibinfo {author} {\bibfnamefont {M.~I.}\ \bibnamefont {Finkel}}, \bibinfo {author} {\bibfnamefont {G.~N.}\ \bibnamefont {Goltsman}}, \bibinfo {author} {\bibfnamefont {F.}~\bibnamefont {Müller}}, \bibinfo {author} {\bibfnamefont {T.}~\bibnamefont {Scheller}},\ and\ \bibinfo {author} {\bibfnamefont {A.~D.}\ \bibnamefont {Semenov}},\ }\href {https://doi.org/10.1088/0953-2048/28/5/055002} {\bibfield  {journal} {\bibinfo  {journal} {Supercond. Sci. Technol.}\ }\textbf {\bibinfo {volume} {28}},\ \bibinfo {pages} {055002} (\bibinfo {year} {2015})}\BibitemShut {NoStop}%
\bibitem [{\citenamefont {Solinas}\ \emph {et~al.}(2015{\natexlab{a}})\citenamefont {Solinas}, \citenamefont {Bosisio},\ and\ \citenamefont {Giazotto}}]{solinas2015extended}%
  \BibitemOpen
  \bibfield  {author} {\bibinfo {author} {\bibfnamefont {P.}~\bibnamefont {Solinas}}, \bibinfo {author} {\bibfnamefont {R.}~\bibnamefont {Bosisio}},\ and\ \bibinfo {author} {\bibfnamefont {F.}~\bibnamefont {Giazotto}},\ }\href {https://doi.org/10.1063/1.4930573} {\bibfield  {journal} {\bibinfo  {journal} {Journal of Applied Physics}\ }\textbf {\bibinfo {volume} {118}},\ \bibinfo {pages} {113901} (\bibinfo {year} {2015}{\natexlab{a}})}\BibitemShut {NoStop}%
\bibitem [{\citenamefont {Solinas}\ \emph {et~al.}(2015{\natexlab{b}})\citenamefont {Solinas}, \citenamefont {Gasparinetti}, \citenamefont {Golubev},\ and\ \citenamefont {Giazotto}}]{solinas2015jrcg}%
  \BibitemOpen
  \bibfield  {author} {\bibinfo {author} {\bibfnamefont {P.}~\bibnamefont {Solinas}}, \bibinfo {author} {\bibfnamefont {S.}~\bibnamefont {Gasparinetti}}, \bibinfo {author} {\bibfnamefont {D.}~\bibnamefont {Golubev}},\ and\ \bibinfo {author} {\bibfnamefont {F.}~\bibnamefont {Giazotto}},\ }\href {https://doi.org/10.1038/srep12260} {\bibfield  {journal} {\bibinfo  {journal} {Scientific Reports}\ }\textbf {\bibinfo {volume} {5}},\ \bibinfo {pages} {12260} (\bibinfo {year} {2015}{\natexlab{b}})}\BibitemShut {NoStop}%
\bibitem [{\citenamefont {Benz}\ and\ \citenamefont {Hamilton}(1998)}]{JAWS}%
  \BibitemOpen
  \bibfield  {author} {\bibinfo {author} {\bibfnamefont {S.~P.}\ \bibnamefont {Benz}}\ and\ \bibinfo {author} {\bibfnamefont {C.~A.}\ \bibnamefont {Hamilton}},\ }\href@noop {} {\bibfield  {journal} {\bibinfo  {journal} {IEEE Trans. Appl. Supercond.}\ }\textbf {\bibinfo {volume} {8}},\ \bibinfo {pages} {42} (\bibinfo {year} {1998})}\BibitemShut {NoStop}%
\bibitem [{\citenamefont {Oakes}\ \emph {et~al.}(2023)\citenamefont {Oakes} \emph {et~al.}}]{fernando_multiplier}%
  \BibitemOpen
  \bibfield  {author} {\bibinfo {author} {\bibfnamefont {G.~A.}\ \bibnamefont {Oakes}} \emph {et~al.},\ }\href@noop {} {\bibfield  {journal} {\bibinfo  {journal} {PRX Quantum}\ }\textbf {\bibinfo {volume} {4}},\ \bibinfo {pages} {020346} (\bibinfo {year} {2023})}\BibitemShut {NoStop}%
\bibitem [{\citenamefont {Lähteenmäki}\ \emph {et~al.}(2013)\citenamefont {Lähteenmäki}, \citenamefont {Paraoanu}, \citenamefont {Hassel},\ and\ \citenamefont {Hakonen}}]{lahteenmaki2013}%
  \BibitemOpen
  \bibfield  {author} {\bibinfo {author} {\bibfnamefont {P.}~\bibnamefont {Lähteenmäki}}, \bibinfo {author} {\bibfnamefont {G.~S.}\ \bibnamefont {Paraoanu}}, \bibinfo {author} {\bibfnamefont {J.}~\bibnamefont {Hassel}},\ and\ \bibinfo {author} {\bibfnamefont {P.~J.}\ \bibnamefont {Hakonen}},\ }\href {https://doi.org/10.1073/pnas.1212705110} {\bibfield  {journal} {\bibinfo  {journal} {Proceedings of the National Academy of Sciences}\ }\textbf {\bibinfo {volume} {110}},\ \bibinfo {pages} {4234} (\bibinfo {year} {2013})},\ \Eprint {https://arxiv.org/abs/https://www.pnas.org/doi/pdf/10.1073/pnas.1212705110} {https://www.pnas.org/doi/pdf/10.1073/pnas.1212705110} \BibitemShut {NoStop}%
\bibitem [{\citenamefont {Yan}\ \emph {et~al.}(2021)\citenamefont {Yan}, \citenamefont {Hassel}, \citenamefont {Vesterinen}, \citenamefont {Zhang}, \citenamefont {Ikonen}, \citenamefont {Groenberg}, \citenamefont {Goetz},\ and\ \citenamefont {Moettoenen}}]{Mottonen2021}%
  \BibitemOpen
  \bibfield  {author} {\bibinfo {author} {\bibfnamefont {C.}~\bibnamefont {Yan}}, \bibinfo {author} {\bibfnamefont {J.}~\bibnamefont {Hassel}}, \bibinfo {author} {\bibfnamefont {V.}~\bibnamefont {Vesterinen}}, \bibinfo {author} {\bibfnamefont {J.}~\bibnamefont {Zhang}}, \bibinfo {author} {\bibfnamefont {J.}~\bibnamefont {Ikonen}}, \bibinfo {author} {\bibfnamefont {L.}~\bibnamefont {Groenberg}}, \bibinfo {author} {\bibfnamefont {J.}~\bibnamefont {Goetz}},\ and\ \bibinfo {author} {\bibfnamefont {M.}~\bibnamefont {Moettoenen}},\ }\href {https://doi.org/10.1038/s41928-021-00680-z} {\bibfield  {journal} {\bibinfo  {journal} {Nature Electronics}\ }\textbf {\bibinfo {volume} {4}},\ \bibinfo {pages} {885} (\bibinfo {year} {2021})}\BibitemShut {NoStop}%
\bibitem [{\citenamefont {Astafiev}\ \emph {et~al.}(2007)\citenamefont {Astafiev}, \citenamefont {Inomata}, \citenamefont {Niskanen}, \citenamefont {Yamamoto}, \citenamefont {Pashkin}, \citenamefont {Nakamura},\ and\ \citenamefont {Tsai}}]{astafiev2007}%
  \BibitemOpen
  \bibfield  {author} {\bibinfo {author} {\bibfnamefont {O.}~\bibnamefont {Astafiev}}, \bibinfo {author} {\bibfnamefont {K.}~\bibnamefont {Inomata}}, \bibinfo {author} {\bibfnamefont {A.~O.}\ \bibnamefont {Niskanen}}, \bibinfo {author} {\bibfnamefont {T.}~\bibnamefont {Yamamoto}}, \bibinfo {author} {\bibfnamefont {Y.~A.}\ \bibnamefont {Pashkin}}, \bibinfo {author} {\bibfnamefont {Y.}~\bibnamefont {Nakamura}},\ and\ \bibinfo {author} {\bibfnamefont {J.~S.}\ \bibnamefont {Tsai}},\ }\href {https://doi.org/10.1038/nature06141} {\bibfield  {journal} {\bibinfo  {journal} {Nature}\ }\textbf {\bibinfo {volume} {449}},\ \bibinfo {pages} {588} (\bibinfo {year} {2007})}\BibitemShut {NoStop}%
\bibitem [{\citenamefont {Cassidy}\ \emph {et~al.}(2017)\citenamefont {Cassidy}, \citenamefont {Bruno}, \citenamefont {Rubbert}, \citenamefont {Irfan}, \citenamefont {Kammhuber}, \citenamefont {Schouten}, \citenamefont {Akhmerov},\ and\ \citenamefont {Kouwenhoven}}]{cassidy}%
  \BibitemOpen
  \bibfield  {author} {\bibinfo {author} {\bibfnamefont {M.~C.}\ \bibnamefont {Cassidy}}, \bibinfo {author} {\bibfnamefont {A.}~\bibnamefont {Bruno}}, \bibinfo {author} {\bibfnamefont {S.}~\bibnamefont {Rubbert}}, \bibinfo {author} {\bibfnamefont {M.}~\bibnamefont {Irfan}}, \bibinfo {author} {\bibfnamefont {J.}~\bibnamefont {Kammhuber}}, \bibinfo {author} {\bibfnamefont {R.~N.}\ \bibnamefont {Schouten}}, \bibinfo {author} {\bibfnamefont {A.~R.}\ \bibnamefont {Akhmerov}},\ and\ \bibinfo {author} {\bibfnamefont {L.~P.}\ \bibnamefont {Kouwenhoven}},\ }\href {https://doi.org/10.1126/science.aah6640} {\bibfield  {journal} {\bibinfo  {journal} {Science}\ }\textbf {\bibinfo {volume} {355}},\ \bibinfo {pages} {939} (\bibinfo {year} {2017})}\BibitemShut {NoStop}%
\bibitem [{\citenamefont {Liu}\ \emph {et~al.}(2015)\citenamefont {Liu}, \citenamefont {Stehlik}, \citenamefont {Eichler}, \citenamefont {Gullans}, \citenamefont {Taylor},\ and\ \citenamefont {Petta}}]{liu2015semiconductor}%
  \BibitemOpen
  \bibfield  {author} {\bibinfo {author} {\bibfnamefont {Y.-Y.}\ \bibnamefont {Liu}}, \bibinfo {author} {\bibfnamefont {J.}~\bibnamefont {Stehlik}}, \bibinfo {author} {\bibfnamefont {C.}~\bibnamefont {Eichler}}, \bibinfo {author} {\bibfnamefont {M.~J.}\ \bibnamefont {Gullans}}, \bibinfo {author} {\bibfnamefont {J.~M.}\ \bibnamefont {Taylor}},\ and\ \bibinfo {author} {\bibfnamefont {J.~R.}\ \bibnamefont {Petta}},\ }\href {https://doi.org/10.1126/science.aaa2501} {\bibfield  {journal} {\bibinfo  {journal} {Science}\ }\textbf {\bibinfo {volume} {347}},\ \bibinfo {pages} {285} (\bibinfo {year} {2015})}\BibitemShut {NoStop}%
\bibitem [{\citenamefont {You}\ and\ \citenamefont {Nori}(2011)}]{nori_review}%
  \BibitemOpen
  \bibfield  {author} {\bibinfo {author} {\bibfnamefont {J.-Q.}\ \bibnamefont {You}}\ and\ \bibinfo {author} {\bibfnamefont {F.}~\bibnamefont {Nori}},\ }\href@noop {} {\bibfield  {journal} {\bibinfo  {journal} {Nature}\ }\textbf {\bibinfo {volume} {474}},\ \bibinfo {pages} {589} (\bibinfo {year} {2011})}\BibitemShut {NoStop}%
\bibitem [{\citenamefont {Greco}\ \emph {et~al.}(2026{\natexlab{a}})\citenamefont {Greco}, \citenamefont {Ballu}, \citenamefont {Giazotto},\ and\ \citenamefont {Crippa}}]{greco2025coherent}%
  \BibitemOpen
  \bibfield  {author} {\bibinfo {author} {\bibfnamefont {A.}~\bibnamefont {Greco}}, \bibinfo {author} {\bibfnamefont {X.}~\bibnamefont {Ballu}}, \bibinfo {author} {\bibfnamefont {F.}~\bibnamefont {Giazotto}},\ and\ \bibinfo {author} {\bibfnamefont {A.}~\bibnamefont {Crippa}},\ }\href@noop {} {\bibfield  {journal} {\bibinfo  {journal} {Nature Communications}\ }\textbf {\bibinfo {volume} {17}},\ \bibinfo {pages} {2972} (\bibinfo {year} {2026}{\natexlab{a}})}\BibitemShut {NoStop}%
\bibitem [{\citenamefont {Greco}\ \emph {et~al.}(2026{\natexlab{b}})\citenamefont {Greco}, \citenamefont {Kaikkonen}, \citenamefont {Chirolli}, \citenamefont {Ronzani}, \citenamefont {Senior}, \citenamefont {Giazotto},\ and\ \citenamefont {Crippa}}]{greco2026spectroscopy}%
  \BibitemOpen
  \bibfield  {author} {\bibinfo {author} {\bibfnamefont {A.}~\bibnamefont {Greco}}, \bibinfo {author} {\bibfnamefont {J.-P.}\ \bibnamefont {Kaikkonen}}, \bibinfo {author} {\bibfnamefont {L.}~\bibnamefont {Chirolli}}, \bibinfo {author} {\bibfnamefont {A.}~\bibnamefont {Ronzani}}, \bibinfo {author} {\bibfnamefont {J.}~\bibnamefont {Senior}}, \bibinfo {author} {\bibfnamefont {F.}~\bibnamefont {Giazotto}},\ and\ \bibinfo {author} {\bibfnamefont {A.}~\bibnamefont {Crippa}},\ }\href {https://arxiv.org/abs/2602.08890} {\bibfield  {journal} {\bibinfo  {journal} {arXiv preprint arXiv:2602.08890}\ } (\bibinfo {year} {2026}{\natexlab{b}})},\ \Eprint {https://arxiv.org/abs/2602.08890} {arXiv:2602.08890 [quant-ph]} \BibitemShut {NoStop}%
\bibitem [{\citenamefont {Babenko}\ \emph {et~al.}(2020)\citenamefont {Babenko}, \citenamefont {Boaventura}, \citenamefont {Flowers-Jacobs}, \citenamefont {Brevik}, \citenamefont {Fox}, \citenamefont {Williams}, \citenamefont {Popovic}, \citenamefont {Dresselhaus},\ and\ \citenamefont {Benz}}]{IEEE_2020}%
  \BibitemOpen
  \bibfield  {author} {\bibinfo {author} {\bibfnamefont {A.~A.}\ \bibnamefont {Babenko}}, \bibinfo {author} {\bibfnamefont {A.~S.}\ \bibnamefont {Boaventura}}, \bibinfo {author} {\bibfnamefont {N.~E.}\ \bibnamefont {Flowers-Jacobs}}, \bibinfo {author} {\bibfnamefont {J.~A.}\ \bibnamefont {Brevik}}, \bibinfo {author} {\bibfnamefont {A.~E.}\ \bibnamefont {Fox}}, \bibinfo {author} {\bibfnamefont {D.~F.}\ \bibnamefont {Williams}}, \bibinfo {author} {\bibfnamefont {Z.}~\bibnamefont {Popovic}}, \bibinfo {author} {\bibfnamefont {P.~D.}\ \bibnamefont {Dresselhaus}},\ and\ \bibinfo {author} {\bibfnamefont {S.~P.}\ \bibnamefont {Benz}},\ }in\ \href {https://doi.org/10.1109/IMS30576.2020.9223811} {\emph {\bibinfo {booktitle} {2020 IEEE/MTT-S International Microwave Symposium (IMS)}}}\ (\bibinfo {address} {Los Angeles, CA, USA},\ \bibinfo {year} {2020})\ pp.\ \bibinfo {pages} {936--939}\BibitemShut {NoStop}%
\bibitem [{\citenamefont {Benz}\ and\ \citenamefont {Booi}(1995)}]{464849}%
  \BibitemOpen
  \bibfield  {author} {\bibinfo {author} {\bibfnamefont {S.}~\bibnamefont {Benz}}\ and\ \bibinfo {author} {\bibfnamefont {P.}~\bibnamefont {Booi}},\ }\href {https://doi.org/10.1109/58.464849} {\bibfield  {journal} {\bibinfo  {journal} {IEEE Trans. Ultrason. Ferroelectr. Freq. Control}\ }\textbf {\bibinfo {volume} {42}},\ \bibinfo {pages} {964} (\bibinfo {year} {1995})}\BibitemShut {NoStop}%
\bibitem [{\citenamefont {Erickson}\ \emph {et~al.}(2014)\citenamefont {Erickson}, \citenamefont {Vissers}, \citenamefont {Sandberg}, \citenamefont {Jefferts},\ and\ \citenamefont {Pappas}}]{Pappas_2014}%
  \BibitemOpen
  \bibfield  {author} {\bibinfo {author} {\bibfnamefont {R.~P.}\ \bibnamefont {Erickson}}, \bibinfo {author} {\bibfnamefont {M.~R.}\ \bibnamefont {Vissers}}, \bibinfo {author} {\bibfnamefont {M.}~\bibnamefont {Sandberg}}, \bibinfo {author} {\bibfnamefont {S.~R.}\ \bibnamefont {Jefferts}},\ and\ \bibinfo {author} {\bibfnamefont {D.~P.}\ \bibnamefont {Pappas}},\ }\href {https://doi.org/10.1103/PhysRevLett.113.187002} {\bibfield  {journal} {\bibinfo  {journal} {Phys. Rev. Lett.}\ }\textbf {\bibinfo {volume} {113}},\ \bibinfo {pages} {187002} (\bibinfo {year} {2014})}\BibitemShut {NoStop}%
\bibitem [{\citenamefont {Wang}\ \emph {et~al.}(2021)\citenamefont {Wang}, \citenamefont {Chen},\ and\ \citenamefont {Li}}]{wang2021}%
  \BibitemOpen
  \bibfield  {author} {\bibinfo {author} {\bibfnamefont {S.-P.}\ \bibnamefont {Wang}}, \bibinfo {author} {\bibfnamefont {Z.}~\bibnamefont {Chen}},\ and\ \bibinfo {author} {\bibfnamefont {T.}~\bibnamefont {Li}},\ }\href {https://doi.org/10.1088/1674-1056/abc2bb} {\bibfield  {journal} {\bibinfo  {journal} {Chinese Physics B}\ }\textbf {\bibinfo {volume} {30}},\ \bibinfo {pages} {048501} (\bibinfo {year} {2021})}\BibitemShut {NoStop}%
\bibitem [{\citenamefont {Shin}\ \emph {et~al.}(2022)\citenamefont {Shin}, \citenamefont {Ryu}, \citenamefont {Miri}, \citenamefont {Shim}, \citenamefont {Choi}, \citenamefont {Alu}, \citenamefont {Suh},\ and\ \citenamefont {Cha}}]{shin2022}%
  \BibitemOpen
  \bibfield  {author} {\bibinfo {author} {\bibfnamefont {J.}~\bibnamefont {Shin}}, \bibinfo {author} {\bibfnamefont {Y.}~\bibnamefont {Ryu}}, \bibinfo {author} {\bibfnamefont {M.-A.}\ \bibnamefont {Miri}}, \bibinfo {author} {\bibfnamefont {S.-B.}\ \bibnamefont {Shim}}, \bibinfo {author} {\bibfnamefont {H.}~\bibnamefont {Choi}}, \bibinfo {author} {\bibfnamefont {A.}~\bibnamefont {Alu}}, \bibinfo {author} {\bibfnamefont {J.}~\bibnamefont {Suh}},\ and\ \bibinfo {author} {\bibfnamefont {J.}~\bibnamefont {Cha}},\ }\href {https://doi.org/10.1021/acs.nanolett.2c01503} {\bibfield  {journal} {\bibinfo  {journal} {Nano Letters}\ }\textbf {\bibinfo {volume} {22}},\ \bibinfo {pages} {5459} (\bibinfo {year} {2022})}\BibitemShut {NoStop}%
\bibitem [{\citenamefont {Wang}\ \emph {et~al.}(2024)\citenamefont {Wang}, \citenamefont {Xu}, \citenamefont {Li}, \citenamefont {Shi}, \citenamefont {Jiang}, \citenamefont {Guo}, \citenamefont {Yue}, \citenamefont {Li}, \citenamefont {Zhang}, \citenamefont {Lyu}, \citenamefont {Pan}, \citenamefont {Deng}, \citenamefont {Dong}, \citenamefont {Tu}, \citenamefont {Dong}, \citenamefont {Cao}, \citenamefont {Zhang}, \citenamefont {Jia}, \citenamefont {Sun}, \citenamefont {Kang}, \citenamefont {Chen}, \citenamefont {Wang}, \citenamefont {Wang},\ and\ \citenamefont {Wu}}]{wang2024}%
  \BibitemOpen
  \bibfield  {author} {\bibinfo {author} {\bibfnamefont {C.-G.}\ \bibnamefont {Wang}}, \bibinfo {author} {\bibfnamefont {W.}~\bibnamefont {Xu}}, \bibinfo {author} {\bibfnamefont {C.}~\bibnamefont {Li}}, \bibinfo {author} {\bibfnamefont {L.}~\bibnamefont {Shi}}, \bibinfo {author} {\bibfnamefont {J.}~\bibnamefont {Jiang}}, \bibinfo {author} {\bibfnamefont {T.}~\bibnamefont {Guo}}, \bibinfo {author} {\bibfnamefont {W.-C.}\ \bibnamefont {Yue}}, \bibinfo {author} {\bibfnamefont {T.}~\bibnamefont {Li}}, \bibinfo {author} {\bibfnamefont {P.}~\bibnamefont {Zhang}}, \bibinfo {author} {\bibfnamefont {Y.-Y.}\ \bibnamefont {Lyu}}, \bibinfo {author} {\bibfnamefont {J.}~\bibnamefont {Pan}}, \bibinfo {author} {\bibfnamefont {X.}~\bibnamefont {Deng}}, \bibinfo {author} {\bibfnamefont {Y.}~\bibnamefont {Dong}}, \bibinfo {author} {\bibfnamefont {X.}~\bibnamefont {Tu}}, \bibinfo {author} {\bibfnamefont {S.}~\bibnamefont {Dong}}, \bibinfo {author} {\bibfnamefont {C.}~\bibnamefont {Cao}}, \bibinfo {author} {\bibfnamefont
  {L.}~\bibnamefont {Zhang}}, \bibinfo {author} {\bibfnamefont {X.}~\bibnamefont {Jia}}, \bibinfo {author} {\bibfnamefont {G.}~\bibnamefont {Sun}}, \bibinfo {author} {\bibfnamefont {L.}~\bibnamefont {Kang}}, \bibinfo {author} {\bibfnamefont {J.}~\bibnamefont {Chen}}, \bibinfo {author} {\bibfnamefont {Y.-L.}\ \bibnamefont {Wang}}, \bibinfo {author} {\bibfnamefont {H.}~\bibnamefont {Wang}},\ and\ \bibinfo {author} {\bibfnamefont {P.}~\bibnamefont {Wu}},\ }\href {https://doi.org/10.1038/s41467-024-48224-1} {\bibfield  {journal} {\bibinfo  {journal} {Nature Communications}\ }\textbf {\bibinfo {volume} {15}},\ \bibinfo {pages} {4009} (\bibinfo {year} {2024})}\BibitemShut {NoStop}%
\bibitem [{\citenamefont {Han}\ \emph {et~al.}(2022)\citenamefont {Han}, \citenamefont {Zou}, \citenamefont {Fu}, \citenamefont {Xu}, \citenamefont {Xu},\ and\ \citenamefont {Tang}}]{acoustic}%
  \BibitemOpen
  \bibfield  {author} {\bibinfo {author} {\bibfnamefont {X.}~\bibnamefont {Han}}, \bibinfo {author} {\bibfnamefont {C.-L.}\ \bibnamefont {Zou}}, \bibinfo {author} {\bibfnamefont {W.}~\bibnamefont {Fu}}, \bibinfo {author} {\bibfnamefont {M.}~\bibnamefont {Xu}}, \bibinfo {author} {\bibfnamefont {Y.}~\bibnamefont {Xu}},\ and\ \bibinfo {author} {\bibfnamefont {H.~X.}\ \bibnamefont {Tang}},\ }\href {https://doi.org/10.1103/PhysRevLett.129.107701} {\bibfield  {journal} {\bibinfo  {journal} {Phys. Rev. Lett.}\ }\textbf {\bibinfo {volume} {129}},\ \bibinfo {pages} {107701} (\bibinfo {year} {2022})}\BibitemShut {NoStop}%
\bibitem [{\citenamefont {Bao}\ \emph {et~al.}(2024)\citenamefont {Bao}, \citenamefont {Li}, \citenamefont {Wang}, \citenamefont {Wang}, \citenamefont {Yang}, \citenamefont {Xiong}, \citenamefont {Song}, \citenamefont {Wu}, \citenamefont {Zhang},\ and\ \citenamefont {Duan}}]{bao2024}%
  \BibitemOpen
  \bibfield  {author} {\bibinfo {author} {\bibfnamefont {Z.}~\bibnamefont {Bao}}, \bibinfo {author} {\bibfnamefont {Y.}~\bibnamefont {Li}}, \bibinfo {author} {\bibfnamefont {Z.}~\bibnamefont {Wang}}, \bibinfo {author} {\bibfnamefont {J.}~\bibnamefont {Wang}}, \bibinfo {author} {\bibfnamefont {J.}~\bibnamefont {Yang}}, \bibinfo {author} {\bibfnamefont {H.}~\bibnamefont {Xiong}}, \bibinfo {author} {\bibfnamefont {Y.}~\bibnamefont {Song}}, \bibinfo {author} {\bibfnamefont {Y.}~\bibnamefont {Wu}}, \bibinfo {author} {\bibfnamefont {H.}~\bibnamefont {Zhang}},\ and\ \bibinfo {author} {\bibfnamefont {L.}~\bibnamefont {Duan}},\ }\href {https://doi.org/10.1038/s41467-024-50333-w} {\bibfield  {journal} {\bibinfo  {journal} {Nature Communications}\ }\textbf {\bibinfo {volume} {15}},\ \bibinfo {pages} {5958} (\bibinfo {year} {2024})}\BibitemShut {NoStop}%
\bibitem [{\citenamefont {Toomey}\ \emph {et~al.}(2018)\citenamefont {Toomey}, \citenamefont {Zhao}, \citenamefont {McCaughan},\ and\ \citenamefont {Berggren}}]{toomey_frequency_2018}%
  \BibitemOpen
  \bibfield  {author} {\bibinfo {author} {\bibfnamefont {E.}~\bibnamefont {Toomey}}, \bibinfo {author} {\bibfnamefont {Q.-Y.}\ \bibnamefont {Zhao}}, \bibinfo {author} {\bibfnamefont {A.~N.}\ \bibnamefont {McCaughan}},\ and\ \bibinfo {author} {\bibfnamefont {K.~K.}\ \bibnamefont {Berggren}},\ }\href {https://doi.org/10.1103/PhysRevApplied.9.064021} {\bibfield  {journal} {\bibinfo  {journal} {Phys. Rev. Appl.}\ }\textbf {\bibinfo {volume} {9}},\ \bibinfo {pages} {064021} (\bibinfo {year} {2018})}\BibitemShut {NoStop}%
\bibitem [{\citenamefont {Toomey}(2017)}]{toomey_microwave_2017}%
  \BibitemOpen
  \bibfield  {author} {\bibinfo {author} {\bibfnamefont {E.}~\bibnamefont {Toomey}},\ }\href@noop {} {\bibinfo {title} {Microwave {Response} of {Nonlinear} {Oscillations} in {Resistively} {Shunted} {Superconducting} {Nanowires}}} (\bibinfo {year} {2017})\BibitemShut {NoStop}%
\bibitem [{\citenamefont {Likharev}\ and\ \citenamefont {Semenov}(1991)}]{likharev1991}%
  \BibitemOpen
  \bibfield  {author} {\bibinfo {author} {\bibfnamefont {K.~K.}\ \bibnamefont {Likharev}}\ and\ \bibinfo {author} {\bibfnamefont {V.~K.}\ \bibnamefont {Semenov}},\ }\href {https://doi.org/10.1109/77.80745} {\bibfield  {journal} {\bibinfo  {journal} {IEEE Transactions on Applied Superconductivity}\ }\textbf {\bibinfo {volume} {1}},\ \bibinfo {pages} {3} (\bibinfo {year} {1991})}\BibitemShut {NoStop}%
\bibitem [{\citenamefont {Toomey}\ \emph {et~al.}(2019)\citenamefont {Toomey}, \citenamefont {Segall},\ and\ \citenamefont {Berggren}}]{toomey_design_2019}%
  \BibitemOpen
  \bibfield  {author} {\bibinfo {author} {\bibfnamefont {E.}~\bibnamefont {Toomey}}, \bibinfo {author} {\bibfnamefont {K.}~\bibnamefont {Segall}},\ and\ \bibinfo {author} {\bibfnamefont {K.~K.}\ \bibnamefont {Berggren}},\ }\href {https://doi.org/10.3389/fnins.2019.00933} {\bibfield  {journal} {\bibinfo  {journal} {Frontiers in Neuroscience}\ }\textbf {\bibinfo {volume} {13}},\ \bibinfo {pages} {933} (\bibinfo {year} {2019})}\BibitemShut {NoStop}%
\bibitem [{\citenamefont {McCaughan}\ and\ \citenamefont {Berggren}(2014)}]{mccaughan_superconducting-nanowire_2014}%
  \BibitemOpen
  \bibfield  {author} {\bibinfo {author} {\bibfnamefont {A.~N.}\ \bibnamefont {McCaughan}}\ and\ \bibinfo {author} {\bibfnamefont {K.~K.}\ \bibnamefont {Berggren}},\ }\href {https://doi.org/10.1021/nl502629y} {\bibfield  {journal} {\bibinfo  {journal} {Nano Letters}\ }\textbf {\bibinfo {volume} {14}},\ \bibinfo {pages} {5748} (\bibinfo {year} {2014})}\BibitemShut {NoStop}%
\bibitem [{\citenamefont {Baghdadi}\ \emph {et~al.}(2020)\citenamefont {Baghdadi}, \citenamefont {Allmaras}, \citenamefont {Butters}, \citenamefont {Dane}, \citenamefont {Iqbal}, \citenamefont {McCaughan}, \citenamefont {Toomey}, \citenamefont {Zhao}, \citenamefont {Kozorezov},\ and\ \citenamefont {Berggren}}]{baghdadi_multilayered_2020}%
  \BibitemOpen
  \bibfield  {author} {\bibinfo {author} {\bibfnamefont {R.}~\bibnamefont {Baghdadi}}, \bibinfo {author} {\bibfnamefont {J.~P.}\ \bibnamefont {Allmaras}}, \bibinfo {author} {\bibfnamefont {B.~A.}\ \bibnamefont {Butters}}, \bibinfo {author} {\bibfnamefont {A.~E.}\ \bibnamefont {Dane}}, \bibinfo {author} {\bibfnamefont {S.}~\bibnamefont {Iqbal}}, \bibinfo {author} {\bibfnamefont {A.~N.}\ \bibnamefont {McCaughan}}, \bibinfo {author} {\bibfnamefont {E.~A.}\ \bibnamefont {Toomey}}, \bibinfo {author} {\bibfnamefont {Q.-Y.}\ \bibnamefont {Zhao}}, \bibinfo {author} {\bibfnamefont {A.~G.}\ \bibnamefont {Kozorezov}},\ and\ \bibinfo {author} {\bibfnamefont {K.~K.}\ \bibnamefont {Berggren}},\ }\href {https://doi.org/10.1103/PhysRevApplied.14.054011} {\bibfield  {journal} {\bibinfo  {journal} {Physical Review Applied}\ }\textbf {\bibinfo {volume} {14}},\ \bibinfo {pages} {054011} (\bibinfo {year} {2020})}\BibitemShut {NoStop}%
\bibitem [{\citenamefont {Kerman}\ \emph {et~al.}(2009)\citenamefont {Kerman}, \citenamefont {Yang}, \citenamefont {Molnar}, \citenamefont {Dauler},\ and\ \citenamefont {Berggren}}]{PhysRevB.79.100509}%
  \BibitemOpen
  \bibfield  {author} {\bibinfo {author} {\bibfnamefont {A.~J.}\ \bibnamefont {Kerman}}, \bibinfo {author} {\bibfnamefont {J.~K.~W.}\ \bibnamefont {Yang}}, \bibinfo {author} {\bibfnamefont {R.~J.}\ \bibnamefont {Molnar}}, \bibinfo {author} {\bibfnamefont {E.~A.}\ \bibnamefont {Dauler}},\ and\ \bibinfo {author} {\bibfnamefont {K.~K.}\ \bibnamefont {Berggren}},\ }\href {https://doi.org/10.1103/PhysRevB.79.100509} {\bibfield  {journal} {\bibinfo  {journal} {Phys. Rev. B}\ }\textbf {\bibinfo {volume} {79}},\ \bibinfo {pages} {100509} (\bibinfo {year} {2009})}\BibitemShut {NoStop}%
\bibitem [{\citenamefont {Trupiano}\ \emph {et~al.}(2025)\citenamefont {Trupiano}, \citenamefont {De~Simoni},\ and\ \citenamefont {Giazotto}}]{Trupiano2024QuasiparticleInjection}%
  \BibitemOpen
  \bibfield  {author} {\bibinfo {author} {\bibfnamefont {G.}~\bibnamefont {Trupiano}}, \bibinfo {author} {\bibfnamefont {G.}~\bibnamefont {De~Simoni}},\ and\ \bibinfo {author} {\bibfnamefont {F.}~\bibnamefont {Giazotto}},\ }\href {https://doi.org/10.1103/PhysRevApplied.23.014046} {\bibfield  {journal} {\bibinfo  {journal} {Phys. Rev. Appl.}\ }\textbf {\bibinfo {volume} {23}},\ \bibinfo {pages} {014046} (\bibinfo {year} {2025})}\BibitemShut {NoStop}%
\bibitem [{\citenamefont {Doh}\ \emph {et~al.}(2005)\citenamefont {Doh}, \citenamefont {van Dam}, \citenamefont {Roest}, \citenamefont {Bakkers}, \citenamefont {Kouwenhoven},\ and\ \citenamefont {Franceschi}}]{doi:10.1126/science.1113523}%
  \BibitemOpen
  \bibfield  {author} {\bibinfo {author} {\bibfnamefont {Y.-J.}\ \bibnamefont {Doh}}, \bibinfo {author} {\bibfnamefont {J.~A.}\ \bibnamefont {van Dam}}, \bibinfo {author} {\bibfnamefont {A.~L.}\ \bibnamefont {Roest}}, \bibinfo {author} {\bibfnamefont {E.~P. A.~M.}\ \bibnamefont {Bakkers}}, \bibinfo {author} {\bibfnamefont {L.~P.}\ \bibnamefont {Kouwenhoven}},\ and\ \bibinfo {author} {\bibfnamefont {S.~D.}\ \bibnamefont {Franceschi}},\ }\href {https://doi.org/10.1126/science.1113523} {\bibfield  {journal} {\bibinfo  {journal} {Science}\ }\textbf {\bibinfo {volume} {309}},\ \bibinfo {pages} {272} (\bibinfo {year} {2005})},\ \Eprint {https://arxiv.org/abs/https://www.science.org/doi/pdf/10.1126/science.1113523} {https://www.science.org/doi/pdf/10.1126/science.1113523} \BibitemShut {NoStop}%
\bibitem [{\citenamefont {Abay}\ \emph {et~al.}(2014)\citenamefont {Abay}, \citenamefont {Persson}, \citenamefont {Nilsson}, \citenamefont {Wu}, \citenamefont {Xu}, \citenamefont {Fogelstr\"om}, \citenamefont {Shumeiko},\ and\ \citenamefont {Delsing}}]{PhysRevB.89.214508}%
  \BibitemOpen
  \bibfield  {author} {\bibinfo {author} {\bibfnamefont {S.}~\bibnamefont {Abay}}, \bibinfo {author} {\bibfnamefont {D.}~\bibnamefont {Persson}}, \bibinfo {author} {\bibfnamefont {H.}~\bibnamefont {Nilsson}}, \bibinfo {author} {\bibfnamefont {F.}~\bibnamefont {Wu}}, \bibinfo {author} {\bibfnamefont {H.~Q.}\ \bibnamefont {Xu}}, \bibinfo {author} {\bibfnamefont {M.}~\bibnamefont {Fogelstr\"om}}, \bibinfo {author} {\bibfnamefont {V.}~\bibnamefont {Shumeiko}},\ and\ \bibinfo {author} {\bibfnamefont {P.}~\bibnamefont {Delsing}},\ }\href {https://doi.org/10.1103/PhysRevB.89.214508} {\bibfield  {journal} {\bibinfo  {journal} {Phys. Rev. B}\ }\textbf {\bibinfo {volume} {89}},\ \bibinfo {pages} {214508} (\bibinfo {year} {2014})}\BibitemShut {NoStop}%
\bibitem [{\citenamefont {Paajaste}\ \emph {et~al.}(2015)\citenamefont {Paajaste}, \citenamefont {Amado}, \citenamefont {Roddaro}, \citenamefont {Bergeret}, \citenamefont {Ercolani}, \citenamefont {Sorba},\ and\ \citenamefont {Giazotto}}]{Paajaste2015}%
  \BibitemOpen
  \bibfield  {author} {\bibinfo {author} {\bibfnamefont {J.}~\bibnamefont {Paajaste}}, \bibinfo {author} {\bibfnamefont {M.}~\bibnamefont {Amado}}, \bibinfo {author} {\bibfnamefont {S.}~\bibnamefont {Roddaro}}, \bibinfo {author} {\bibfnamefont {F.~S.}\ \bibnamefont {Bergeret}}, \bibinfo {author} {\bibfnamefont {D.}~\bibnamefont {Ercolani}}, \bibinfo {author} {\bibfnamefont {L.}~\bibnamefont {Sorba}},\ and\ \bibinfo {author} {\bibfnamefont {F.}~\bibnamefont {Giazotto}},\ }\href {https://doi.org/10.1021/nl504544s} {\bibfield  {journal} {\bibinfo  {journal} {Nano Letters}\ }\textbf {\bibinfo {volume} {15}},\ \bibinfo {pages} {1803} (\bibinfo {year} {2015})}\BibitemShut {NoStop}%
\bibitem [{\citenamefont {Kousar}\ \emph {et~al.}(2022)\citenamefont {Kousar}, \citenamefont {Carrad}, \citenamefont {Stampfer}, \citenamefont {Krogstrup}, \citenamefont {Nyg{\aa}rd},\ and\ \citenamefont {Jespersen}}]{Kousar2022}%
  \BibitemOpen
  \bibfield  {author} {\bibinfo {author} {\bibfnamefont {B.}~\bibnamefont {Kousar}}, \bibinfo {author} {\bibfnamefont {D.~J.}\ \bibnamefont {Carrad}}, \bibinfo {author} {\bibfnamefont {L.}~\bibnamefont {Stampfer}}, \bibinfo {author} {\bibfnamefont {P.}~\bibnamefont {Krogstrup}}, \bibinfo {author} {\bibfnamefont {J.}~\bibnamefont {Nyg{\aa}rd}},\ and\ \bibinfo {author} {\bibfnamefont {T.~S.}\ \bibnamefont {Jespersen}},\ }\href {https://doi.org/10.1021/acs.nanolett.2c02532} {\bibfield  {journal} {\bibinfo  {journal} {Nano Letters}\ }\textbf {\bibinfo {volume} {22}},\ \bibinfo {pages} {8845} (\bibinfo {year} {2022})}\BibitemShut {NoStop}%
\bibitem [{\citenamefont {Roddaro}\ \emph {et~al.}(2011)\citenamefont {Roddaro}, \citenamefont {Pescaglini}, \citenamefont {Ercolani}, \citenamefont {Sorba}, \citenamefont {Giazotto},\ and\ \citenamefont {Beltram}}]{roddaro2011hot}%
  \BibitemOpen
  \bibfield  {author} {\bibinfo {author} {\bibfnamefont {S.}~\bibnamefont {Roddaro}}, \bibinfo {author} {\bibfnamefont {A.}~\bibnamefont {Pescaglini}}, \bibinfo {author} {\bibfnamefont {D.}~\bibnamefont {Ercolani}}, \bibinfo {author} {\bibfnamefont {L.}~\bibnamefont {Sorba}}, \bibinfo {author} {\bibfnamefont {F.}~\bibnamefont {Giazotto}},\ and\ \bibinfo {author} {\bibfnamefont {F.}~\bibnamefont {Beltram}},\ }\href@noop {} {\bibfield  {journal} {\bibinfo  {journal} {Nano Research}\ }\textbf {\bibinfo {volume} {4}},\ \bibinfo {pages} {259} (\bibinfo {year} {2011})}\BibitemShut {NoStop}%
\bibitem [{\citenamefont {Spathis}\ \emph {et~al.}(2011)\citenamefont {Spathis}, \citenamefont {Biswas}, \citenamefont {Roddaro}, \citenamefont {Sorba}, \citenamefont {Giazotto},\ and\ \citenamefont {Beltram}}]{spathis2011hybrid}%
  \BibitemOpen
  \bibfield  {author} {\bibinfo {author} {\bibfnamefont {P.}~\bibnamefont {Spathis}}, \bibinfo {author} {\bibfnamefont {S.}~\bibnamefont {Biswas}}, \bibinfo {author} {\bibfnamefont {S.}~\bibnamefont {Roddaro}}, \bibinfo {author} {\bibfnamefont {L.}~\bibnamefont {Sorba}}, \bibinfo {author} {\bibfnamefont {F.}~\bibnamefont {Giazotto}},\ and\ \bibinfo {author} {\bibfnamefont {F.}~\bibnamefont {Beltram}},\ }\href@noop {} {\bibfield  {journal} {\bibinfo  {journal} {Nanotechnology}\ }\textbf {\bibinfo {volume} {22}},\ \bibinfo {pages} {105201} (\bibinfo {year} {2011})}\BibitemShut {NoStop}%
\bibitem [{\citenamefont {Strambini}\ \emph {et~al.}(2020)\citenamefont {Strambini}, \citenamefont {Iorio}, \citenamefont {Durante}, \citenamefont {Citro}, \citenamefont {Sanz-Fern{\'a}ndez}, \citenamefont {Guarcello}, \citenamefont {Tokatly}, \citenamefont {Braggio}, \citenamefont {Rocci}, \citenamefont {Ligato} \emph {et~al.}}]{strambini2020josephson}%
  \BibitemOpen
  \bibfield  {author} {\bibinfo {author} {\bibfnamefont {E.}~\bibnamefont {Strambini}}, \bibinfo {author} {\bibfnamefont {A.}~\bibnamefont {Iorio}}, \bibinfo {author} {\bibfnamefont {O.}~\bibnamefont {Durante}}, \bibinfo {author} {\bibfnamefont {R.}~\bibnamefont {Citro}}, \bibinfo {author} {\bibfnamefont {C.}~\bibnamefont {Sanz-Fern{\'a}ndez}}, \bibinfo {author} {\bibfnamefont {C.}~\bibnamefont {Guarcello}}, \bibinfo {author} {\bibfnamefont {I.~V.}\ \bibnamefont {Tokatly}}, \bibinfo {author} {\bibfnamefont {A.}~\bibnamefont {Braggio}}, \bibinfo {author} {\bibfnamefont {M.}~\bibnamefont {Rocci}}, \bibinfo {author} {\bibfnamefont {N.}~\bibnamefont {Ligato}}, \emph {et~al.},\ }\href@noop {} {\bibfield  {journal} {\bibinfo  {journal} {Nature Nanotechnology}\ }\textbf {\bibinfo {volume} {15}},\ \bibinfo {pages} {656} (\bibinfo {year} {2020})}\BibitemShut {NoStop}%
\bibitem [{\citenamefont {Iorio}\ \emph {et~al.}(2018)\citenamefont {Iorio}, \citenamefont {Rocci}, \citenamefont {Bours}, \citenamefont {Carrega}, \citenamefont {Zannier}, \citenamefont {Sorba}, \citenamefont {Roddaro}, \citenamefont {Giazotto},\ and\ \citenamefont {Strambini}}]{iorio2018vectorial}%
  \BibitemOpen
  \bibfield  {author} {\bibinfo {author} {\bibfnamefont {A.}~\bibnamefont {Iorio}}, \bibinfo {author} {\bibfnamefont {M.}~\bibnamefont {Rocci}}, \bibinfo {author} {\bibfnamefont {L.}~\bibnamefont {Bours}}, \bibinfo {author} {\bibfnamefont {M.}~\bibnamefont {Carrega}}, \bibinfo {author} {\bibfnamefont {V.}~\bibnamefont {Zannier}}, \bibinfo {author} {\bibfnamefont {L.}~\bibnamefont {Sorba}}, \bibinfo {author} {\bibfnamefont {S.}~\bibnamefont {Roddaro}}, \bibinfo {author} {\bibfnamefont {F.}~\bibnamefont {Giazotto}},\ and\ \bibinfo {author} {\bibfnamefont {E.}~\bibnamefont {Strambini}},\ }\href@noop {} {\bibfield  {journal} {\bibinfo  {journal} {Nano letters}\ }\textbf {\bibinfo {volume} {19}},\ \bibinfo {pages} {652} (\bibinfo {year} {2018})}\BibitemShut {NoStop}%
\bibitem [{\citenamefont {Heersche}\ \emph {et~al.}(2007)\citenamefont {Heersche}, \citenamefont {Jarillo-Herrero}, \citenamefont {Oostinga}, \citenamefont {Vandersypen},\ and\ \citenamefont {Morpurgo}}]{Heersche2007}%
  \BibitemOpen
  \bibfield  {author} {\bibinfo {author} {\bibfnamefont {H.~B.}\ \bibnamefont {Heersche}}, \bibinfo {author} {\bibfnamefont {P.}~\bibnamefont {Jarillo-Herrero}}, \bibinfo {author} {\bibfnamefont {J.~B.}\ \bibnamefont {Oostinga}}, \bibinfo {author} {\bibfnamefont {L.~M.~K.}\ \bibnamefont {Vandersypen}},\ and\ \bibinfo {author} {\bibfnamefont {A.~F.}\ \bibnamefont {Morpurgo}},\ }\href {https://doi.org/10.1038/nature05555} {\bibfield  {journal} {\bibinfo  {journal} {Nature}\ }\textbf {\bibinfo {volume} {446}},\ \bibinfo {pages} {56} (\bibinfo {year} {2007})}\BibitemShut {NoStop}%
\bibitem [{\citenamefont {Nguyen}\ \emph {et~al.}(1990)\citenamefont {Nguyen}, \citenamefont {Werking}, \citenamefont {Kroemer},\ and\ \citenamefont {Hu}}]{10.1063/1.103546}%
  \BibitemOpen
  \bibfield  {author} {\bibinfo {author} {\bibfnamefont {C.}~\bibnamefont {Nguyen}}, \bibinfo {author} {\bibfnamefont {J.}~\bibnamefont {Werking}}, \bibinfo {author} {\bibfnamefont {H.}~\bibnamefont {Kroemer}},\ and\ \bibinfo {author} {\bibfnamefont {E.~L.}\ \bibnamefont {Hu}},\ }\href {https://doi.org/10.1063/1.103546} {\bibfield  {journal} {\bibinfo  {journal} {Applied Physics Letters}\ }\textbf {\bibinfo {volume} {57}},\ \bibinfo {pages} {87} (\bibinfo {year} {1990})}\BibitemShut {NoStop}%
\bibitem [{\citenamefont {S\"ut\ifmmode~\mbox{\H{o}}\else \H{o}\fi{}}\ \emph {et~al.}(2022)\citenamefont {S\"ut\ifmmode~\mbox{\H{o}}\else \H{o}\fi{}}, \citenamefont {Prok}, \citenamefont {Makk}, \citenamefont {Kirti}, \citenamefont {Biasiol}, \citenamefont {Csonka},\ and\ \citenamefont {T\'ov\'ari}}]{PhysRevB.106.235404}%
  \BibitemOpen
  \bibfield  {author} {\bibinfo {author} {\bibfnamefont {M.}~\bibnamefont {S\"ut\ifmmode~\mbox{\H{o}}\else \H{o}\fi{}}}, \bibinfo {author} {\bibfnamefont {T.}~\bibnamefont {Prok}}, \bibinfo {author} {\bibfnamefont {P.}~\bibnamefont {Makk}}, \bibinfo {author} {\bibfnamefont {M.}~\bibnamefont {Kirti}}, \bibinfo {author} {\bibfnamefont {G.}~\bibnamefont {Biasiol}}, \bibinfo {author} {\bibfnamefont {S.}~\bibnamefont {Csonka}},\ and\ \bibinfo {author} {\bibfnamefont {E.}~\bibnamefont {T\'ov\'ari}},\ }\href {https://doi.org/10.1103/PhysRevB.106.235404} {\bibfield  {journal} {\bibinfo  {journal} {Phys. Rev. B}\ }\textbf {\bibinfo {volume} {106}},\ \bibinfo {pages} {235404} (\bibinfo {year} {2022})}\BibitemShut {NoStop}%
\bibitem [{\citenamefont {Paghi}\ \emph {et~al.}(2025{\natexlab{a}})\citenamefont {Paghi}, \citenamefont {Trupiano}, \citenamefont {Simoni}, \citenamefont {Arif}, \citenamefont {Sorba},\ and\ \citenamefont {Giazotto}}]{Paghi2025}%
  \BibitemOpen
  \bibfield  {author} {\bibinfo {author} {\bibfnamefont {A.}~\bibnamefont {Paghi}}, \bibinfo {author} {\bibfnamefont {G.}~\bibnamefont {Trupiano}}, \bibinfo {author} {\bibfnamefont {G.~D.}\ \bibnamefont {Simoni}}, \bibinfo {author} {\bibfnamefont {O.}~\bibnamefont {Arif}}, \bibinfo {author} {\bibfnamefont {L.}~\bibnamefont {Sorba}},\ and\ \bibinfo {author} {\bibfnamefont {F.}~\bibnamefont {Giazotto}},\ }\href {https://doi.org/10.1002/adfm.202416957} {\bibfield  {journal} {\bibinfo  {journal} {Advanced Functional Materials}\ }\textbf {\bibinfo {volume} {35}},\ \bibinfo {pages} {2416957} (\bibinfo {year} {2025}{\natexlab{a}})}\BibitemShut {NoStop}%
\bibitem [{\citenamefont {Battisti}\ \emph {et~al.}(2024)\citenamefont {Battisti}, \citenamefont {De~Simoni}, \citenamefont {Braggio}, \citenamefont {Paghi}, \citenamefont {Sorba},\ and\ \citenamefont {Giazotto}}]{10.1063/5.0225361}%
  \BibitemOpen
  \bibfield  {author} {\bibinfo {author} {\bibfnamefont {S.}~\bibnamefont {Battisti}}, \bibinfo {author} {\bibfnamefont {G.}~\bibnamefont {De~Simoni}}, \bibinfo {author} {\bibfnamefont {A.}~\bibnamefont {Braggio}}, \bibinfo {author} {\bibfnamefont {A.}~\bibnamefont {Paghi}}, \bibinfo {author} {\bibfnamefont {L.}~\bibnamefont {Sorba}},\ and\ \bibinfo {author} {\bibfnamefont {F.}~\bibnamefont {Giazotto}},\ }\href {https://doi.org/10.1063/5.0225361} {\bibfield  {journal} {\bibinfo  {journal} {Applied Physics Letters}\ }\textbf {\bibinfo {volume} {125}},\ \bibinfo {pages} {202601} (\bibinfo {year} {2024})}\BibitemShut {NoStop}%
\bibitem [{\citenamefont {Paghi}\ \emph {et~al.}(2025{\natexlab{b}})\citenamefont {Paghi}, \citenamefont {Borgongino}, \citenamefont {Battisti}, \citenamefont {Tortorella}, \citenamefont {Trupiano}, \citenamefont {Simoni}, \citenamefont {Strambini}, \citenamefont {Sorba},\ and\ \citenamefont {Giazotto}}]{Paghi2025a}%
  \BibitemOpen
  \bibfield  {author} {\bibinfo {author} {\bibfnamefont {A.}~\bibnamefont {Paghi}}, \bibinfo {author} {\bibfnamefont {L.}~\bibnamefont {Borgongino}}, \bibinfo {author} {\bibfnamefont {S.}~\bibnamefont {Battisti}}, \bibinfo {author} {\bibfnamefont {S.}~\bibnamefont {Tortorella}}, \bibinfo {author} {\bibfnamefont {G.}~\bibnamefont {Trupiano}}, \bibinfo {author} {\bibfnamefont {G.~D.}\ \bibnamefont {Simoni}}, \bibinfo {author} {\bibfnamefont {E.}~\bibnamefont {Strambini}}, \bibinfo {author} {\bibfnamefont {L.}~\bibnamefont {Sorba}},\ and\ \bibinfo {author} {\bibfnamefont {F.}~\bibnamefont {Giazotto}},\ }\href {https://doi.org/10.1021/acsaelm.5c00038} {\bibfield  {journal} {\bibinfo  {journal} {ACS Applied Electronic Materials}\ }\textbf {\bibinfo {volume} {7}},\ \bibinfo {pages} {3756} (\bibinfo {year} {2025}{\natexlab{b}})}\BibitemShut {NoStop}%
\bibitem [{\citenamefont {Paghi}\ \emph {et~al.}(2025{\natexlab{c}})\citenamefont {Paghi}, \citenamefont {Borgongino}, \citenamefont {Tortorella}, \citenamefont {Simoni}, \citenamefont {Strambini}, \citenamefont {Sorba},\ and\ \citenamefont {Giazotto}}]{Paghi2025b}%
  \BibitemOpen
  \bibfield  {author} {\bibinfo {author} {\bibfnamefont {A.}~\bibnamefont {Paghi}}, \bibinfo {author} {\bibfnamefont {L.}~\bibnamefont {Borgongino}}, \bibinfo {author} {\bibfnamefont {S.}~\bibnamefont {Tortorella}}, \bibinfo {author} {\bibfnamefont {G.~D.}\ \bibnamefont {Simoni}}, \bibinfo {author} {\bibfnamefont {E.}~\bibnamefont {Strambini}}, \bibinfo {author} {\bibfnamefont {L.}~\bibnamefont {Sorba}},\ and\ \bibinfo {author} {\bibfnamefont {F.}~\bibnamefont {Giazotto}},\ }\href {https://doi.org/10.1038/s41467-025-62931-3} {\bibfield  {journal} {\bibinfo  {journal} {Nature Communications}\ }\textbf {\bibinfo {volume} {16}},\ \bibinfo {pages} {8442} (\bibinfo {year} {2025}{\natexlab{c}})}\BibitemShut {NoStop}%
\bibitem [{\citenamefont {Capotondi}\ \emph {et~al.}(2004)\citenamefont {Capotondi}, \citenamefont {Biasiol}, \citenamefont {Vobornik}, \citenamefont {Sorba}, \citenamefont {Giazotto}, \citenamefont {Cavallini},\ and\ \citenamefont {Fraboni}}]{capotondi2004two}%
  \BibitemOpen
  \bibfield  {author} {\bibinfo {author} {\bibfnamefont {F.}~\bibnamefont {Capotondi}}, \bibinfo {author} {\bibfnamefont {G.}~\bibnamefont {Biasiol}}, \bibinfo {author} {\bibfnamefont {I.}~\bibnamefont {Vobornik}}, \bibinfo {author} {\bibfnamefont {L.}~\bibnamefont {Sorba}}, \bibinfo {author} {\bibfnamefont {F.}~\bibnamefont {Giazotto}}, \bibinfo {author} {\bibfnamefont {A.}~\bibnamefont {Cavallini}},\ and\ \bibinfo {author} {\bibfnamefont {B.}~\bibnamefont {Fraboni}},\ }\href@noop {} {\bibfield  {journal} {\bibinfo  {journal} {Journal of Vacuum Science \& Technology B: Microelectronics and Nanometer Structures Processing, Measurement, and Phenomena}\ }\textbf {\bibinfo {volume} {22}},\ \bibinfo {pages} {702} (\bibinfo {year} {2004})}\BibitemShut {NoStop}%
\bibitem [{\citenamefont {Desrat}\ \emph {et~al.}(2004)\citenamefont {Desrat}, \citenamefont {Giazotto}, \citenamefont {Pellegrini}, \citenamefont {Beltram}, \citenamefont {Capotondi}, \citenamefont {Biasiol}, \citenamefont {Sorba},\ and\ \citenamefont {Maude}}]{desrat2004magnetotransport}%
  \BibitemOpen
  \bibfield  {author} {\bibinfo {author} {\bibfnamefont {W.}~\bibnamefont {Desrat}}, \bibinfo {author} {\bibfnamefont {F.}~\bibnamefont {Giazotto}}, \bibinfo {author} {\bibfnamefont {V.}~\bibnamefont {Pellegrini}}, \bibinfo {author} {\bibfnamefont {F.}~\bibnamefont {Beltram}}, \bibinfo {author} {\bibfnamefont {F.}~\bibnamefont {Capotondi}}, \bibinfo {author} {\bibfnamefont {G.}~\bibnamefont {Biasiol}}, \bibinfo {author} {\bibfnamefont {L.}~\bibnamefont {Sorba}},\ and\ \bibinfo {author} {\bibfnamefont {D.}~\bibnamefont {Maude}},\ }\href@noop {} {\bibfield  {journal} {\bibinfo  {journal} {Physical Review B—Condensed Matter and Materials Physics}\ }\textbf {\bibinfo {volume} {69}},\ \bibinfo {pages} {245324} (\bibinfo {year} {2004})}\BibitemShut {NoStop}%
\bibitem [{\citenamefont {Kawakami}\ and\ \citenamefont {Takayanagi}(1985)}]{10.1063/1.95809}%
  \BibitemOpen
  \bibfield  {author} {\bibinfo {author} {\bibfnamefont {T.}~\bibnamefont {Kawakami}}\ and\ \bibinfo {author} {\bibfnamefont {H.}~\bibnamefont {Takayanagi}},\ }\href {https://doi.org/10.1063/1.95809} {\bibfield  {journal} {\bibinfo  {journal} {Applied Physics Letters}\ }\textbf {\bibinfo {volume} {46}},\ \bibinfo {pages} {92} (\bibinfo {year} {1985})}\BibitemShut {NoStop}%
\bibitem [{\citenamefont {Chrestin}\ and\ \citenamefont {Merkt}(1997)}]{10.1063/1.119116}%
  \BibitemOpen
  \bibfield  {author} {\bibinfo {author} {\bibfnamefont {A.}~\bibnamefont {Chrestin}}\ and\ \bibinfo {author} {\bibfnamefont {U.}~\bibnamefont {Merkt}},\ }\href {https://doi.org/10.1063/1.119116} {\bibfield  {journal} {\bibinfo  {journal} {Applied Physics Letters}\ }\textbf {\bibinfo {volume} {70}},\ \bibinfo {pages} {3149} (\bibinfo {year} {1997})}\BibitemShut {NoStop}%
\bibitem [{\citenamefont {Mayer}\ \emph {et~al.}(2020)\citenamefont {Mayer}, \citenamefont {Schiela}, \citenamefont {Yuan}, \citenamefont {Hatefipour}, \citenamefont {Sarney}, \citenamefont {Svensson}, \citenamefont {Leff}, \citenamefont {Campos}, \citenamefont {Wickramasinghe}, \citenamefont {Dartiailh}, \citenamefont {{\v{Z}}uti{\'c}},\ and\ \citenamefont {Shabani}}]{Mayer2020}%
  \BibitemOpen
  \bibfield  {author} {\bibinfo {author} {\bibfnamefont {W.}~\bibnamefont {Mayer}}, \bibinfo {author} {\bibfnamefont {W.~F.}\ \bibnamefont {Schiela}}, \bibinfo {author} {\bibfnamefont {J.}~\bibnamefont {Yuan}}, \bibinfo {author} {\bibfnamefont {M.}~\bibnamefont {Hatefipour}}, \bibinfo {author} {\bibfnamefont {W.~L.}\ \bibnamefont {Sarney}}, \bibinfo {author} {\bibfnamefont {S.~P.}\ \bibnamefont {Svensson}}, \bibinfo {author} {\bibfnamefont {A.~C.}\ \bibnamefont {Leff}}, \bibinfo {author} {\bibfnamefont {T.}~\bibnamefont {Campos}}, \bibinfo {author} {\bibfnamefont {K.~S.}\ \bibnamefont {Wickramasinghe}}, \bibinfo {author} {\bibfnamefont {M.~C.}\ \bibnamefont {Dartiailh}}, \bibinfo {author} {\bibfnamefont {I.}~\bibnamefont {{\v{Z}}uti{\'c}}},\ and\ \bibinfo {author} {\bibfnamefont {J.}~\bibnamefont {Shabani}},\ }\href {https://doi.org/10.1021/acsaelm.0c00269} {\bibfield  {journal} {\bibinfo  {journal} {ACS Applied Electronic Materials}\ }\textbf {\bibinfo {volume} {2}},\ \bibinfo {pages} {2351} (\bibinfo {year}
  {2020})}\BibitemShut {NoStop}%
\bibitem [{\citenamefont {Amado}\ \emph {et~al.}(2013)\citenamefont {Amado}, \citenamefont {Fornieri}, \citenamefont {Carillo}, \citenamefont {Biasiol}, \citenamefont {Sorba}, \citenamefont {Pellegrini},\ and\ \citenamefont {Giazotto}}]{amado2013electrostatic}%
  \BibitemOpen
  \bibfield  {author} {\bibinfo {author} {\bibfnamefont {M.}~\bibnamefont {Amado}}, \bibinfo {author} {\bibfnamefont {A.}~\bibnamefont {Fornieri}}, \bibinfo {author} {\bibfnamefont {F.}~\bibnamefont {Carillo}}, \bibinfo {author} {\bibfnamefont {G.}~\bibnamefont {Biasiol}}, \bibinfo {author} {\bibfnamefont {L.}~\bibnamefont {Sorba}}, \bibinfo {author} {\bibfnamefont {V.}~\bibnamefont {Pellegrini}},\ and\ \bibinfo {author} {\bibfnamefont {F.}~\bibnamefont {Giazotto}},\ }\href@noop {} {\bibfield  {journal} {\bibinfo  {journal} {Physical Review B—Condensed Matter and Materials Physics}\ }\textbf {\bibinfo {volume} {87}},\ \bibinfo {pages} {134506} (\bibinfo {year} {2013})}\BibitemShut {NoStop}%
\bibitem [{\citenamefont {Amado}\ \emph {et~al.}(2014)\citenamefont {Amado}, \citenamefont {Fornieri}, \citenamefont {Biasiol}, \citenamefont {Sorba},\ and\ \citenamefont {Giazotto}}]{amado2014ballistic}%
  \BibitemOpen
  \bibfield  {author} {\bibinfo {author} {\bibfnamefont {M.}~\bibnamefont {Amado}}, \bibinfo {author} {\bibfnamefont {A.}~\bibnamefont {Fornieri}}, \bibinfo {author} {\bibfnamefont {G.}~\bibnamefont {Biasiol}}, \bibinfo {author} {\bibfnamefont {L.}~\bibnamefont {Sorba}},\ and\ \bibinfo {author} {\bibfnamefont {F.}~\bibnamefont {Giazotto}},\ }\href@noop {} {\bibfield  {journal} {\bibinfo  {journal} {Applied Physics Letters}\ }\textbf {\bibinfo {volume} {104}} (\bibinfo {year} {2014})}\BibitemShut {NoStop}%
\bibitem [{\citenamefont {Andreev}(1964)}]{Andreev1964}%
  \BibitemOpen
  \bibfield  {author} {\bibinfo {author} {\bibfnamefont {A.~F.}\ \bibnamefont {Andreev}},\ }\href@noop {} {\bibfield  {journal} {\bibinfo  {journal} {Soviet Physics JETP}\ }\textbf {\bibinfo {volume} {19}},\ \bibinfo {pages} {1228} (\bibinfo {year} {1964})}\BibitemShut {NoStop}%
\bibitem [{\citenamefont {Pannetier}\ and\ \citenamefont {Courtois}(2000)}]{Pannetier2000}%
  \BibitemOpen
  \bibfield  {author} {\bibinfo {author} {\bibfnamefont {B.}~\bibnamefont {Pannetier}}\ and\ \bibinfo {author} {\bibfnamefont {H.}~\bibnamefont {Courtois}},\ }\href {https://doi.org/10.1023/A:1004631231633} {\bibfield  {journal} {\bibinfo  {journal} {Journal of Low Temperature Physics}\ }\textbf {\bibinfo {volume} {118}},\ \bibinfo {pages} {599} (\bibinfo {year} {2000})}\BibitemShut {NoStop}%
\bibitem [{\citenamefont {Golubov}\ \emph {et~al.}(2004)\citenamefont {Golubov}, \citenamefont {Kupriyanov},\ and\ \citenamefont {Il'ichev}}]{RevModPhys.76.411}%
  \BibitemOpen
  \bibfield  {author} {\bibinfo {author} {\bibfnamefont {A.~A.}\ \bibnamefont {Golubov}}, \bibinfo {author} {\bibfnamefont {M.~Y.}\ \bibnamefont {Kupriyanov}},\ and\ \bibinfo {author} {\bibfnamefont {E.}~\bibnamefont {Il'ichev}},\ }\href {https://doi.org/10.1103/RevModPhys.76.411} {\bibfield  {journal} {\bibinfo  {journal} {Rev. Mod. Phys.}\ }\textbf {\bibinfo {volume} {76}},\ \bibinfo {pages} {411} (\bibinfo {year} {2004})}\BibitemShut {NoStop}%
\bibitem [{\citenamefont {Beenakker}\ and\ \citenamefont {van Houten}(1991)}]{PhysRevLett.66.3056}%
  \BibitemOpen
  \bibfield  {author} {\bibinfo {author} {\bibfnamefont {C.~W.~J.}\ \bibnamefont {Beenakker}}\ and\ \bibinfo {author} {\bibfnamefont {H.}~\bibnamefont {van Houten}},\ }\href {https://doi.org/10.1103/PhysRevLett.66.3056} {\bibfield  {journal} {\bibinfo  {journal} {Phys. Rev. Lett.}\ }\textbf {\bibinfo {volume} {66}},\ \bibinfo {pages} {3056} (\bibinfo {year} {1991})}\BibitemShut {NoStop}%
\bibitem [{\citenamefont {Wen}\ \emph {et~al.}(2019)\citenamefont {Wen}, \citenamefont {Shabani},\ and\ \citenamefont {Tutuc}}]{8915971}%
  \BibitemOpen
  \bibfield  {author} {\bibinfo {author} {\bibfnamefont {F.}~\bibnamefont {Wen}}, \bibinfo {author} {\bibfnamefont {J.}~\bibnamefont {Shabani}},\ and\ \bibinfo {author} {\bibfnamefont {E.}~\bibnamefont {Tutuc}},\ }\href {https://doi.org/10.1109/TED.2019.2951634} {\bibfield  {journal} {\bibinfo  {journal} {IEEE Transactions on Electron Devices}\ }\textbf {\bibinfo {volume} {66}},\ \bibinfo {pages} {5367} (\bibinfo {year} {2019})}\BibitemShut {NoStop}%
\bibitem [{\citenamefont {Tinkham}(1996)}]{Tinkham1996}%
  \BibitemOpen
  \bibfield  {author} {\bibinfo {author} {\bibfnamefont {M.}~\bibnamefont {Tinkham}},\ }\href@noop {} {\emph {\bibinfo {title} {Introduction to Superconductivity}}},\ \bibinfo {edition} {2nd}\ ed.\ (\bibinfo  {publisher} {McGraw-Hill},\ \bibinfo {address} {New York},\ \bibinfo {year} {1996})\BibitemShut {NoStop}%
\bibitem [{\citenamefont {{Analog Devices}}()}]{LTspice}%
  \BibitemOpen
  \bibfield  {author} {\bibinfo {author} {\bibnamefont {{Analog Devices}}},\ }\href {https://www.analog.com/ltspice} {\bibinfo {title} {{LTspice XVII}}},\ \bibinfo {note} {circuit simulation software}\BibitemShut {NoStop}%
\bibitem [{\citenamefont {Kiviranta}(2021)}]{Kiviranta2021}%
  \BibitemOpen
  \bibfield  {author} {\bibinfo {author} {\bibfnamefont {M.}~\bibnamefont {Kiviranta}},\ }\href@noop {} {\bibfield  {journal} {\bibinfo  {journal} {arXiv:2103.11465}\ } (\bibinfo {year} {2021})}\BibitemShut {NoStop}%
\bibitem [{\citenamefont {Barone}\ and\ \citenamefont {Patern\`o}(1982)}]{BaronePaterno1982}%
  \BibitemOpen
  \bibfield  {author} {\bibinfo {author} {\bibfnamefont {A.}~\bibnamefont {Barone}}\ and\ \bibinfo {author} {\bibfnamefont {G.}~\bibnamefont {Patern\`o}},\ }\href@noop {} {\emph {\bibinfo {title} {Physics and Applications of the Josephson Effect}}}\ (\bibinfo  {publisher} {Wiley},\ \bibinfo {address} {New York},\ \bibinfo {year} {1982})\BibitemShut {NoStop}%
\bibitem [{\citenamefont {Hao}\ \emph {et~al.}(2024)\citenamefont {Hao}, \citenamefont {Zhao}, \citenamefont {Huang}, \citenamefont {Deng}, \citenamefont {Yang}, \citenamefont {Ru}, \citenamefont {Liu}, \citenamefont {Wan}, \citenamefont {Liu}, \citenamefont {Li}, \citenamefont {Wang}, \citenamefont {Tu}, \citenamefont {Zhang}, \citenamefont {Jia}, \citenamefont {Wu}, \citenamefont {Chen}, \citenamefont {Kang},\ and\ \citenamefont {Wu}}]{Hao2024}%
  \BibitemOpen
  \bibfield  {author} {\bibinfo {author} {\bibfnamefont {H.}~\bibnamefont {Hao}}, \bibinfo {author} {\bibfnamefont {Q.-Y.}\ \bibnamefont {Zhao}}, \bibinfo {author} {\bibfnamefont {Y.-H.}\ \bibnamefont {Huang}}, \bibinfo {author} {\bibfnamefont {J.}~\bibnamefont {Deng}}, \bibinfo {author} {\bibfnamefont {F.}~\bibnamefont {Yang}}, \bibinfo {author} {\bibfnamefont {S.-Y.}\ \bibnamefont {Ru}}, \bibinfo {author} {\bibfnamefont {Z.}~\bibnamefont {Liu}}, \bibinfo {author} {\bibfnamefont {C.}~\bibnamefont {Wan}}, \bibinfo {author} {\bibfnamefont {H.}~\bibnamefont {Liu}}, \bibinfo {author} {\bibfnamefont {Z.-J.}\ \bibnamefont {Li}}, \bibinfo {author} {\bibfnamefont {H.-B.}\ \bibnamefont {Wang}}, \bibinfo {author} {\bibfnamefont {X.-C.}\ \bibnamefont {Tu}}, \bibinfo {author} {\bibfnamefont {L.-B.}\ \bibnamefont {Zhang}}, \bibinfo {author} {\bibfnamefont {X.-Q.}\ \bibnamefont {Jia}}, \bibinfo {author} {\bibfnamefont {X.-L.}\ \bibnamefont {Wu}}, \bibinfo {author} {\bibfnamefont {J.}~\bibnamefont {Chen}}, \bibinfo
  {author} {\bibfnamefont {L.}~\bibnamefont {Kang}},\ and\ \bibinfo {author} {\bibfnamefont {P.-H.}\ \bibnamefont {Wu}},\ }\href {https://doi.org/10.1038/s41377-023-01374-1} {\bibfield  {journal} {\bibinfo  {journal} {Light: Science \& Applications}\ }\textbf {\bibinfo {volume} {13}},\ \bibinfo {pages} {25} (\bibinfo {year} {2024})}\BibitemShut {NoStop}%
\end{thebibliography}

%

\end{document}